\documentclass[aps,prd,twocolumn,showpacs,nofootinbib]{revtex4-1}

\usepackage{graphicx}
\usepackage{amsmath}
\usepackage{amssymb}
\usepackage{bm}
\usepackage{bbm}
\usepackage{color}

\begin{document}

\title{Exclusive charmonium production from $e^+ e^-$ annihilation round the $Z^0$ peak}

\author{Gu Chen}
\author{Xing-Gang Wu}
\email{email:wuxg@cqu.edu.cn}
\author{Zhan Sun}
\author{Sheng-Quan Wang}
\author{Jian-Ming Shen}

\address{Department of Physics, Chongqing University, Chongqing 401331, P.R. China}

\date{\today}

\begin{abstract}
We make a comparative and comprehensive study on the charmonium exclusive productions at the $e^+e^-$ collider with the collision energy either round the $Z^0$-boson mass for a super $Z$ factory or equals to $10.6$ GeV for the $B$ factories as Belle and BABAR. We study the total cross sections for the charmonium production via the exclusive processes $e^+e^- \to \gamma^*/Z^0 \to H_{1}+H_{2}$ and $e^+e^- \to \gamma^*/Z^0 \to H_{1} +\gamma$, where $H_{1}$ and $H_{2}$ represent the dominant color-singlet $S$-wave and $P$-wave charmonium states respectively. Total cross sections versus the $e^+e^-$ collision energy $\sqrt{s}$, together with their uncertainties, are presented, which clearly show the relative importance of these channels. At the $B$ factory, the production channels via the virtual $\gamma^*$ propagator are dominant over the channels via the $Z^0$ propagator by about four orders. While, at the super $Z$ factory, due to the $Z^0$-boson resonance effect, the $Z^0$ boson channels shall provide sizable or even dominant contributions in comparison to the channels via the $\gamma^*$ propagator. Sizable exclusive charmonium events can be produced at the super $Z$ factory with high luminocity up to $10^{36}{\rm cm}^{-2}{\rm s}^{-1}$, especially for the channel of $e^+e^- \to Z^0 \to H_{1} +\gamma$, e.g. by taking $m_c=1.50\pm0.20$ GeV, we shall have $(5.0^{+0.8}_{-0.6})\times10^4$ $J/\psi$, $(7.5^{+1.1}_{-0.9})\times10^3$ $\eta_c$, $(6.2^{+3.3}_{-1.9})\times10^3$ $h_{c}$, $(3.1^{+1.7}_{-0.9})\times10^2$ $\chi_{c0}$, $(2.2^{+1.0}_{-0.4})\times10^3$ $\chi_{c1}$, and $(7.7^{+4.1}_{-2.4})\times10^2$ $\chi_{c2}$ events by one operation year. Thus, in addition to the $B$ factories as BABAR and Belle, such a super $Z$ factory shall provide another useful platform for studying the heavy quarkonium properties and for testing QCD theories.
\end{abstract}


\maketitle

\section{Introduction}

The heavy quarkonium provides an ideal platform to investigate the properties of the bound states. Considering the fact of the nonrelativistic nature of heavy quark or antiquark inside the quarkonium, the nonrelativistic QCD (NRQCD)~\cite{nrqcd} provides a powerful tool in studying the production mechanism of heavy quarkonium. The NRQCD is an effective field theory for separating the relativistic effects from the nonrelativistic contributions of different $v_{Q}$ order, $v_{Q} (\ll 1)$ being the typical velocity of heavy quark or antiquark in the quarkonium rest frame, whose relative importance can be estimated by the velocity scaling rules. It provides the definitions of nonperturbative contributions from the long-distance part and makes us possible to do numerical calculations directly. In the present paper, we shall concentrate our attention on the charmonium exclusive production at the $e^+ e^-$ collider. That is, we shall compute the cross-sections for the exclusive processes $e^+e^- \to H_{1}+H_{2}$ and $e^+e^- \to H_{1} +\gamma$, where $H_{1}$ and $H_{2}$ represent the dominant color-singlet $S$-wave and $P$-wave charmonium states respectively. All discussions shall be based on the NRQCD framework and the color singlet mechanism~\cite{npb172425}.

The double charmonium exclusive production channel $e^{+}e^{-} \to J/\psi+\eta_{c}$ has been measured by the Belle~\cite{Teva,Tevb} and the BABAR collaborations~\cite{BABAR}. The measured cross section is $$\sigma[e^{+}e^{-} \to J/\psi+\eta_{c}]\times {\mathcal B}^{\eta_{c}}[\geq2]=25.6\pm2.8\pm3.4 \;{\rm fb}$$ for Belle collaboration~\cite{Teva,Tevb} and $$\sigma[e^{+}e^{-}\to J/\psi+\eta_{c}] \times {\mathcal B}^{\eta_{c}}[\geq2]=17.6\pm2.8\pm2.1 \;{\rm fb}$$ for BABAR collaboration~\cite{BABAR}, in which the first error stands for the statistic error and the second error stands for the systematic error. Here ${\mathcal B}^{\eta_{c}}[\geq2]$ denotes the branching fraction for the $\eta_c$ meson decaying into at least two charged tracks. These measurements are unexpectedly large in comparison with the leading order (LO) calculation~\cite{BLB2003,BL2003,plB557(2003)45-54}. These discrepancies between experimental measurements and theoretical predictions are challenging issues in NRQCD. It arouses people's great interest, many suggestions have been tried to explain the puzzle~\cite{prd67054007,prd77014002,sum rule,Braguta2008,plb57039,NLO corrections, relativistic, BLC2008, Wang20081,Wang20082,wang2012,wang2013}, in which either the higher $\alpha_s$ order pQCD corrections or the relativistic corrections or the nonperturbative corrections have been considered.

By taking both the radiative and relativistic corrections into account, Ref.\cite{BLC2008} got $\sigma[e^{+}e^{-} \to J/\psi+\eta_{c}] = 17.6^{+8.1}_{-6.7} \; {\rm fb}$, the authors there then optimistically concluded that such disagreement has been resolved. However, one may doubt the validity of the $\alpha_s$- and $v^2$- expansion in the process $e^{+}+e^{-}\rightarrow J/\psi+\eta_c$, if the LO result is an order of magnitude smaller than the experimental measurements and the next-to-leading order (NLO) corrections/higher $v^2$-expansion terms inversely play a dominate role for the double charmonium production. Furthermore, Refs.\cite{Wang20081,Wang20082} showed that the large renormalization scale dependence of the cross section under the conventional scale setting can not be improved even with the NLO correction \footnote{It is noted that such renormalization scale dependence can be solved by using the newly suggested principle of maximum conformality~\cite{pmc} even at the NLO level, c.f. Ref.~\cite{wangpmc}.}. It is therefore helpful to find another experimental platform to check all theoretical estimations. A super $Z$ factory running at an energy around the $Z^0$-boson mass with a high luminosity ${\cal L}\simeq 10^{34-36}{\rm cm}^{-2}{\rm s}^{-1}$~\cite{zfactory}, similar to the GigaZ program of the Internal Linear Collider~\cite{ILC1,ILC2}, can be a useful reference for experimental studies, complementing the present BaBar and Belle results on heavy quarkonia.

The $e^{+}e^{-} \to {\rm charmonium} + \gamma$ is another important channel for studying the heavy quarkonium physics~\cite{single-charm1,single-charm,prdDL,higher-charm,higher-charm2}. It has been roughly estimated~\cite{single-charm} that at the $B$ factory, the total cross section of $e^+e^-\to H+\gamma$ can be greater than that of $e^+e^-\to H + J/\psi$ by about two orders of magnitude for $H=\eta_c$, and by about $1{\rm -}10$ times for $H$ equals to the spin-triplet $P$-wave charmonium state. Therefore, if the background from the channel $e^+e^-\to X+\gamma$ is under well control in the recoil mass ($m_X$) region near the $H$ resonance, the process $e^+e^-\to H+\gamma$ can be detected by analyzing the photon energy spectrum in $e^+e^-\to X+\gamma$. Because only one bound state in the final state, one can adopt it to extract more subtle properties of the charmonium. Considering the high luminosity and a clean environment at the super $Z$ factory~\cite{zfactory}, due to the $Z^0$-boson resonance effect, more and more rare decays and productions can be observed and measured. It is interesting to show how the charmonium is produced at the super $Z$ factory.

In this paper, we shall make a comparative and comprehensive study on those two types of processes $e^+e^-\to \gamma^*/Z^{0}\to H_{1} +H_{2}$ and $e^+e^-\to \gamma^*/Z^{0} \to H+\gamma$ at the super $Z$ factory and the $B$ factory. All calculations are done at the leading $\alpha_s$ order but with dominant relative corrections being included. Because the $c$-quark line is massive, the squared amplitude of process is too complex and lengthy, especially for the $P$-wave case. To improve the calculation efficiency, we adopt the improved trace technology~\cite{itt0,zbc0,tbc,zbc1,zbc2,eebc,wbc} to deal with the hard scattering amplitude directly at the amplitude level, which can simplify the amplitude as much as possible. As an explanation of the approach, we first arrange the amplitude $M_{ss^{\prime}}$ into four orthogonal sub-amplitudes $M_{i}$ according to the spins of the ingoing electron with spin $s$ and the positron with spin $s'$, then transform these sub-amplitudes into a trace form. Then we do the trace of the Dirac $\gamma$-matrix strings at the amplitude level, which finally result in analytic series over some (limited) independent Lorentz-structures.

The remaining parts of the paper are organized as follows. In Sec.II, we present the calculation technology for the processes $e^+e^-\to \gamma^*/Z^{0}\to H_{1}+H_{2}$ and $e^+e^-\to \gamma^*/Z^{0} \to H+\gamma$. Numerical results and discussions for the total cross sections and the corresponding differential distributions are presented in Sec.III. Sec.IV is reserved for a summary. Useful formulas for the processes are given in the Appendix.

\section{Calculation technology}

\begin{figure}[htb]
\includegraphics[width=0.49\textwidth]{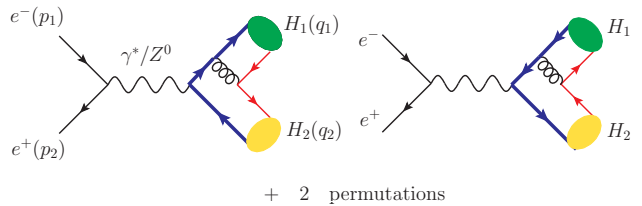}
\caption{Feynman diagrams for the process $e^+(p_{2})+e^-(p_{1})\to \gamma^*/Z^{0} \to H_{1}(q_{1})+H_{2}(q_{2})$ at the tree level, $H_1$ or $H_2$ stands for color-singlet charmonium states: $|[c\bar{c}]_{\bf 1}(^1S_0)\rangle$, $|[c\bar{c}]_{\bf 1}(^3S_1)\rangle$, $|[c\bar{c}]_{\bf 1}(^1P_1)\rangle$, and $|[c\bar{c}]_{\bf 1}(^3P_J)\rangle$ ($J=0,1,2$), respectively. The other two permutation diagrams are obtained by exchanging the position of the thicker and thinner $c$-quark lines.  } \label{fig1}
\end{figure}

\begin{figure}[htb]
\includegraphics[width=0.49\textwidth]{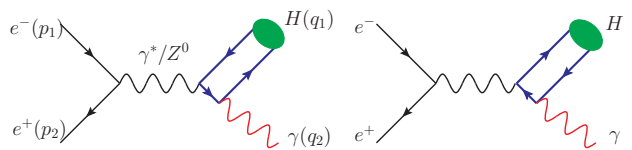}
\caption{ Feynman diagrams for the process $e^+(p_{2})+e^-(p_{1})\to \gamma^{*} / Z^{0} \to H(q_{1})+\gamma(q_{2})$ at the tree level, where $H$ stands for the color-singlet $S$-wave and $P$-wave charmonium states: $H=|[c\bar{c}]_{\bf 1}(^1S_0)\rangle$, $|[c\bar{c}]_{\bf 1}(^3S_1)\rangle$, $|[c\bar{c}]_{\bf 1}(^1P_1)\rangle$, and $|[c\bar{c}]_{\bf 1}(^3P_J)\rangle$ ($J=0,1,2$), respectively. } \label{fig2}
\end{figure}

Typical Feynman diagrams for the processes $e^+(p_{2})+e^-(p_{1})\to \gamma^*/Z^{0} \to H_{1}(q_{1})+H_{2}(q_{2})$ and $e^+(p_{2})+e^-(p_{1})\to \gamma^*/Z^{0} \to H(q_{1})+\gamma(q_{2})$ are presented in Figs.(\ref{fig1},\ref{fig2}). The treatment of both processes are similar. In the following, we take the process $e^+ e^- \to \gamma^*/Z^{0} \to H+\gamma$ as an example to illustrate our approach. We put all the necessary basic Lorentz structures together with their coefficients for these two processes in the Appendix.

In the present paper, we shall concentrate our attention on the dominant color-singlet charmonium states $H=|(c\bar{c})_{\bf 1}[n]\rangle$ with $n={}^{2S+1}L_J $ that corresponds to ${}^1S_0$, ${}^3S_1$, ${}^1P_1$ and ${}^3P_J$ ($J$=0, 1, 2) states, respectively. The subscript ${\bf (1)}$ means the $(c\bar{c})$-pair is in color-singlet state. The symbols $S$, $L$ and $J$ are quantum numbers for the spin angular momentum, the orbital angular momentum and the total angular momentum of the charmonium, respectively. We will not take the color-octet states into consideration, since it is noted that due to the color-suppression for the hard scattering amplitude and also the suppression from the non-perturbative color-octet matrix element, the color-octet charmonium states shall give negligible contributions to those processes. This is somewhat different from the charmonium semi-exclusive processes, where the color-octet states indeed provide sizable contributions~\cite{sun}.

\subsection{Cross sections for the processes}

Within the NRQCD framework, the cross section for the process $e^+(p_{2})+e^-(p_{1})\to \gamma^*/Z^{0} \to H(q_{1})+\gamma(q_{2})$ can be factorized as
\begin{equation}
d\sigma=\sum_{n} d\hat\sigma(e^+e^- \to (c\bar{c})[n]+\gamma) \langle{\cal O}^H(n) \rangle \;,
\end{equation}
where $\langle{\cal O}^H(n) \rangle$ is the non-perturbative but universal matrix element which represents the hadronization probability of the perturbative state $(c\bar{c})[n]$ into the bound state. The color-singlet matrix elements can be directly related to the wave functions at the origin for the $S$-wave state or the first derivative of the wave function at the origin for the $P$-wave state accordingly, which can be computed via the potential models and/or potential NRQCD and/or lattice QCD, respectively.

The differential cross section $d\hat\sigma(e^+e^- \to (c\bar{c})[n]+\gamma)$ stands for the $2\to2$ short-distance cross section, i.e.
\begin{eqnarray}
&& d\hat\sigma(e^{+}+e^{-}\rightarrow (c\bar{c})[n]+\gamma)  \nonumber\\
&& = \frac{1}{4\sqrt{(p_1\cdot p_2)^2-m_{e^+}^2 m_{e^-}^2}} \overline{\sum}  |{\cal M}|^{2} d\Phi_2 \;.
\end{eqnarray}
The symbol $\overline{\sum}$ means we need to average over the initial degrees of freedom and sum over the final ones. The two-body phase space is defined as,
\begin{equation}
d{\Phi _2} = {(2\pi )^4} {\delta ^4}({p_1} + {p_2} - \sum\limits_f^2 {{q_f}} )\prod\limits_{f = 1}^2 {\frac{d\vec{q}_{f}} {(2\pi)^3 2q_f^0}} .
\end{equation}
After doing the integration over the $\delta$-function and the azimuth angle, we obtain, $$\int d{\Phi _2} = \frac{-|\vec{q}_{1}|} {8 \pi \sqrt{s}} \cdot d(\cos \theta ) \;,$$ which is in the center-of-mass frame. The parameter $s$ stands for the squared center-of-mass energy. The magnitude of the charmonium momentum $|\vec{q}_1| = {(s -M_H^2)}/{2\sqrt{s}}$, and $\theta$ is the angle between $\vec{p}_1$ and $\vec{q}_1$.

\subsection{Hard scattering amplitude}

The hard scattering amplitude ${\cal M}$ can be written in the following form,
\begin{equation}
i \;{\cal M} = {\cal C}\;\sum\limits_{\kappa} {{{\bar{v}}_{s'}}(p_2) L^{\mu} u_{s}(p_1)} {\cal A}_{\kappa}^{\nu} D_{\mu\nu} , \label{amplitude}
\end{equation}
where $\kappa$ denotes the number of independent Feynman diagrams for a given process. The overall color factor ${\cal C}$ equals to $\frac{4}{3}$ or $\sqrt{3}$ for the case of double charmonium production or for one charmonium production respectively. The subscripts $s$ and $s'$ represent the spin projections of the initial particles. For the process via the virtual photon, we have
\begin{subequations}
\begin{eqnarray}
{L^\mu } =  - ie{\gamma ^\mu },\\
D_{\mu\nu}=\frac{-ig_{\mu\nu}}{k^2},
\end{eqnarray}
\end{subequations}
and for the process via the $Z^0$ boson propagator, we have
\begin{subequations}
\begin{eqnarray}
{L^\mu } = \frac{{i g_w}} {{4\cos {\theta _w}}}{\gamma ^\mu }(1 - 4{\sin ^2}{\theta _w} - {\gamma ^5}) \;, \\
D_{\mu\nu}=\frac{i}{k^2-m^2_Z +im_Z\Gamma_z}\left(-g_{\mu\nu}+{k_\mu k_\nu}/{k^2}\right),
\end{eqnarray}
\end{subequations}
where $\Gamma_z$ stands for the total decay width of the $Z^0$ boson. The parameter $e$ is the unit of the electric charge and $g_w$ is the weak interaction coupling constant. If the charmonium is in $S$-wave state ($L = 0$) with $S = 0$ or $1$, the strings of the Dirac $\gamma$-matrices ${\cal A}_{n}^{\nu}$ are
\begin{widetext}
\begin{eqnarray}
{\cal A}_1^{\nu (S = 0,L = 0)} &=& i \; \textrm{Tr} \left.\left[ {\Pi _{{q_1}}^0(q)\not\!\varepsilon ({q_2})\frac{{{\not\!q}_3} + {m_c}}{{q_3^2 - m_c^2}}{L^\nu }} \right]\right|_{q = 0} \;, \\
{\cal A}_2^{\nu (S = 0,L = 0)} &=& i \; \textrm{Tr} \left.\left[ {\Pi _{{q_1}}^0(q){L^\nu }\frac{{-{{\not\!q}_3} + {m_c}}}
{{q_3^2 - m_c^2}}\not\!\varepsilon ({q_2})} \right]\right|_{q = 0}\;, \\
{\cal A}_1^{\nu (S = 1,L = 0)} &=& i \; \textrm{Tr} \left. \left[\varepsilon_{s,\beta}(q_1) {\Pi _{{q_1}}^{\beta}(q)\not\! \varepsilon ({q_2})\frac{{{{\not\!q}_3} + {m_c}}} {{q_3^2 - m_c^2}}{L^\nu }}\right] \right|_{q = 0} \;, \\
{\cal A}_2^{\nu (S = 1,L = 0)} &=& i \; \textrm{Tr} \left. \left[\varepsilon_{s,\beta}(q_1) {\Pi _{{q_1}}^{\beta}(q){L^\nu }\frac{{-{{\not\!q}_3} + {m_c}}} {{q_3^2 - m_c^2}}\not\!\varepsilon ({q_2})} \right] \right|_{q = 0}\;.
\end{eqnarray}
\end{widetext}

If the charmonium is in $P$-wave states ($L = 1$) with $S = 0$ or $1$, the strings of the Dirac $\gamma$-matrices ${\cal A}_{n}^{\nu}$ are
\begin{widetext}
\begin{eqnarray}
{\cal A}_1^{\nu (S = 0,L = 1)} &=& i \; \frac{d}{dq_{\alpha}} \textrm{Tr} \left.\left[\varepsilon_{l,\alpha}(q_1){\Pi _{{q_1}}^0(q) \not\!\varepsilon ({q_2})\frac{{{{\not\!q}_3} + {m_c}}} {{q_3^2 - m_c^2}}{L^\nu }} \right] \right|_{q = 0}\;, \\
{\cal A}_2^{\nu (S = 0,L = 1)} &=& i \; \frac{d}{dq_{\alpha}} \textrm{Tr}\left.\left[\varepsilon_{l,\alpha}(q_1){\Pi _{{q_1}}^0(q){L^\nu }\frac{{-{{\not\!q}_3} + {m_c}}} {{q_3^2 - m_c^2}}\not\!\varepsilon ({q_2})} \right]\right|_{q = 0}\;, \\
{\cal A}_1^{\nu (S = 1,L = 1)} &=& i \; \frac{d}{dq_{\alpha}} \textrm{Tr}\left.\left[\varepsilon^{J}_{\alpha\beta}(q_1){\Pi _{{q_1}}^{\beta}(q) \not\!\varepsilon({q_2})\frac{{{{\not\!q}_3} + {m_c}}}
{{q_3^2 - m_c^2}}{L^\nu }} \right]\right|_{q = 0}\;, \\
{\cal A}_2^{\nu (S = 1,L = 1)} &=& i \; \frac{d}{dq_{\alpha}} \textrm{Tr}\left.\left[\varepsilon^{J}_{\alpha\beta}(q_1) {\Pi _{{q_1}}^{\beta}(q){L^\nu }\frac{{-{{\not\!q}_3} + {m_c}}} {{q_3^2 - m_c^2}}\not\!\varepsilon ({q_2})} \right]\right|_{q = 0} \;,
\end{eqnarray}
\end{widetext}
where $q_{3}=\frac{q_1} {2} + {q_2} + q$ and $q$ is the relative momentum  between the constitute quarks of the charmonium. The covariant form of the projectors are~\cite{proj0,proj1,proj2,proj3},
\begin{displaymath}
\Pi _{{q_1}}^0(q)=\frac{1} {{\sqrt {8m_c^3} }} \left(\frac{{{{\not\!q}_1}}}
{2} - {\not\!q } - {m_c}\right){\gamma ^5} \left(\frac{{{{\not\!q}_1}}}
{2} + {\not\!q } + {m_c}\right)
\end{displaymath}
and
\begin{displaymath}
\Pi _{{q_1}}^{\beta}(q)=\frac{1} {{\sqrt {8m_c^3} }} \left(\frac{{{{\not\!q}_1}}} {2} - {\not\!q } - {m_c}\right) {\gamma ^\beta } \left(\frac{{{{\not\!q}_1}}} {2} + {\not\!q } + {m_c}\right)\;.
\end{displaymath}
The projectors by including the relativistic effect can be found in Ref.\cite{proj4}. Such relativistic effect only provides small effect to the hard part of the process, so at present, we adopt the above conventionally adopted projectors to do our calculation. The sum over the polarization for a spin-triplet $S$-wave state ($^3S_1$) or a spin-singlet $P$-wave state ($^1P_1$) is given by,
\begin{equation}
\Pi_{\alpha\beta} =\sum_{J_z}\varepsilon_\alpha \varepsilon^*_{\beta} =-g_{\alpha\beta}+\frac{q_{1\alpha} q_{1\beta}}{q_{1}^2}\;,
\end{equation}
where $\varepsilon$ stands for the polarization vector $\varepsilon_l$ or $\varepsilon_s$ respectively. The sum over the polarization for the spin-triplet $P$-wave states ($^3P_J$ with $J=0,1,2$) is given by~\cite{proj2,proj3},
\begin{eqnarray}
\varepsilon^{(0)}_{\alpha\beta} \varepsilon^{(0)*}_{\alpha'\beta'} &=& \frac{1}{3} \Pi_{\alpha\beta}\Pi_{\alpha'\beta'}, \\
\sum_{J_z}\varepsilon^{(1)}_{\alpha\beta} \varepsilon^{(1)*}_{\alpha'\beta'} &=& \frac{1}{2}(\Pi_{\alpha\alpha'}\Pi_{\beta\beta'}- \Pi_{\alpha\beta'}\Pi_{\alpha'\beta}), \\
\sum_{J_z}\varepsilon^{(2)}_{\alpha\beta} \varepsilon^{(2)*}_{\alpha'\beta'} &=& \frac{1}{2}({\Pi_{\alpha\alpha'}\Pi_{\beta\beta'}+ \Pi_{\alpha\beta'}\Pi_{\alpha'\beta}}) -\frac{1}{3}\Pi_{\alpha\beta}\Pi_{\alpha'\beta'}\;.
\end{eqnarray}

\subsection{Simplified amplitude under the improved trace technology}

We take the improved trace technology to deal with the amplitude ${\cal M}$. Detailed processes of the approach can be found in Refs.\cite{itt0,zbc0,tbc,zbc1,zbc2,eebc,wbc}, for self-consistency, we shall present its main idea and our main results here.

For the purpose, we introduce the negative and positive helicity states for a massless spinor $u(k_0)$, which satisfy
\begin{equation} \label{refu}
u_{\pm}(k_0)\bar{u}_{\pm}(k_0)=\omega_{\pm} \not\!{k}_0 \;,
\end{equation}
where $\omega_{\pm}=(1\pm\gamma_5)/2$. Treating the negative helicity spinor $u_{-}(k_0)$ as the reference input, and the positive helicity spinor $u_{+}(k_0)$ can be redefined by introducing an arbitrary space-like momentum $k_1$ as
\begin{equation}
u_{+}(k_0)=\not\!{k}_1 u_{-}(k_0) .
\end{equation}

All massive spinors $u_s(q)$ and $v_{s'}(q)$ with momentum $q$ can be constructed by the massless spinors $u(k_0)$ with explicit positive or negative helicity as,
\begin{eqnarray}
u_{\pm s}(q)&=&(\not\!{q}+m)u_{\mp}(k_0)/\sqrt{2k_0\cdot q}\;, \\
v_{\pm s'}(q)&=&(\not\!{q}-m)u_{\mp}(k_0)/\sqrt{2k_0\cdot q}\;.
\end{eqnarray}
For convenience, we set the hard scattering amplitude ${\cal M}=\left(M_{ss'}+M_{-s-s'}+M_{-ss'}+M_{s-s'}\right)$. After transforming all the spinors into $u(k_0)$, we can write down the amplitude $M_{\pm s\pm s'}$ with four possible spin projections in a trace form with the help of Eq.(\ref{refu}):
\begin{eqnarray}
M_{ss'} &=& N \textrm{Tr}[(\not\!{p}_1+m_e)\cdot \omega_{-}\not\!{k_0}\cdot(\not\!{p}_2-m_e)\cdot A]\;, \nonumber\\
M_{-s-s'} &=& N \textrm{Tr}[(\not\!{p}_1+m_e)\cdot \omega_{+}\not\!{k_0}\cdot(\not\!{p}_2-m_e)\cdot A]\;, \nonumber\\
M_{-ss'} &=& N \textrm{Tr}[(\not\!{p}_1+m_e)\cdot \omega_{-}\not\!{k_0}\not\!{k}_1\cdot(\not\!{p_2}-m_e)\cdot A]\;,\nonumber\\
M_{s-s'} &=& N \textrm{Tr}[(\not\!{p}_1+m_e)\cdot\omega_{+}\not\!{k_1} \not\!{k_0}\cdot(\not\!{p}_2-m_e)\cdot A]\;, \nonumber
\end{eqnarray}
where the overall factor $N={{\cal C} } /\sqrt{4(k_0\cdot p_1) (k_0\cdot p_2)}$ and $A=\sum\limits_{n = 1}^{2} L^{\mu}{\cal A}_{n}^{\nu}D_{\mu\nu}$. For simplifying the calculation, we rearrange $M_{ss'}$ into $M_{n}$ ($n=1,$$ \cdots $$,4$)~\cite{itt0,zbc0,tbc,zbc1,zbc2,eebc,wbc}:
\begin{eqnarray}
M_1 &=& L_1 \textrm{Tr}[(\not\!{p_1}+m_e)\cdot(\not\!{p_2}-m_e)\cdot A]\;,\\
M_2 &=& L_2 \textrm{Tr}[(\not\!{p_1}+m_e)\cdot\gamma_5\cdot(\not\!{p_2}-m_e)\cdot A] \;,\\
\label{m3}
M_3 &=& -L_2 \textrm{Tr}[(\not\!{p_1}+m_e)\cdot\not\!{k_1}\cdot(\not\!{p_2}-m_e)\cdot A] \;,\\
\label{m4}
M_4 &=& L_1 \textrm{Tr}[(\not\!{p_1}+m_e)\cdot\gamma_5\cdot\not\!{k_1}\cdot(\not\!{p_2}-m_e)\cdot A]\;,
\end{eqnarray}
where
\begin{displaymath}
L_{1}=\frac{{\cal C}}{2\sqrt{p_1\cdot p_2-m_e^2}}\;\;{\rm and}\;\;
L_{2}=\frac{{\cal C}}{2\sqrt{p_1\cdot p_2+m_e^2}} \;.
\end{displaymath}
It's easily to check that,
\begin{equation}
|{\cal M}|^{2} = |M_{1}|^2 + |M_{2}|^2 + |M_{3}|^2 + |M_{4}|^2.
\end{equation}

Finally, we expand the amplitudes $M_{n}$ over the independent Lorentz structures as
\begin{equation}
M_{n}=\sum^{\eta}_{j=1} A^{n}_j B_j \;\; (n=1-4),
\end{equation}
The parameter $\eta$ is the maximum number of basic Lorentz structures $B_{j}$ for specific production channels. Generally, the values of $\eta$, $B_{j}$ and $A^{n}_j$ are different for different channels. We put all the necessary Lorentz structures together with their non-zero coefficients of the present channels in the appendix.

Due to more particles, more massive quark lines, or more loops for higher-order calculation are involved in high energy processes, the amplitude of the process becomes more and more complex and lengthy. The improved trace technology is very helpful for improving the generating efficiency in comparison to the conventional squared amplitude approach, many of its applications have already been done in the literature. It could be a tedious task to find out all the independent $B_{j}$ and all the non-zero $A^{n}_j$, especially for the $P$-wave case and for those processes involving more particles in the final states. For convenience, we suggest a \emph{Mathematica} program based on the Feyncalc package~\cite{feyncalc} to automatically and effectively extract the independent Lorentz structures together with their coefficients from $M_{n}$, which are available upon request.

As a cross-check, in addition to the improved trace technology, we also adopt the traditional trace technology for dealing with the mentioned processes. Numerically, we obtain a good agreement between these two approaches within reasonable numerical errors for all the mentioned channels.

\section{Numerical results}

\subsection{Input parameters}

In calculating the hard scattering amplitude of the process, the charmonium mass $M_{H(c\bar{c})}$ is taken as $2m_c$, which ensures the gauge invariance of the hard scattering amplitude. We take the effective charm quark mass $m_c=1.5$ GeV as its central value throughout the paper, which is consistent with the $1S$ scheme for setting the heavy quark masses~\cite{1scheme}. While, by taking the relativistic effect into consideration, one will obtain a phase-space factor $\left(1 - M_{H_{c\bar{c}}}^2 / s\right)$~\cite{single-charm}, which shows such difference may lead to a sizable correction at the order of ${\cal O}(m_{c} v^2)$. In this phase-space factor, we take the charmonium masses as~\cite{pdg}: $M_{J/\psi}=3.097$ GeV, $M_{\eta_{c}}=2.980$ GeV, $M_{\chi_{c0}}=3.415$ GeV, $M_{\chi_{c1}}=3.511$ GeV and $M_{\chi_{c2}}=3.556$ GeV, respectively.

Other input parameters are taken as~\cite{pdg}: $m_Z=91.1876$ GeV, $\Gamma_z=2.4952$ GeV, ${\sin ^2}{\theta_w}=0.2312$, $\alpha_s(m_Z)=0.1184$, the fine-structure constant $\alpha={1}/{130.9}$, and the weak interaction coupling constant $g_{w}={e}/{\sin\theta_{w}}$. The renormalization scale is set to be $2m_c$ for charmonium and $\alpha_s(2m_c)=0.26$ for leading order $\alpha_s$. The color-singlet non-perturbative matrix elements are related to the wave function at the origin $|\psi_s(0)|={|R_s(0)|}/{\sqrt{4\pi}}$ for $S$-wave state and its first derivative at the origin $|\psi'_p(0)|=\sqrt{\frac{3}{4\pi}}|R'_p(0)|$ for $P$-wave state, respectively. These matrix elements can be calculated by using the potential model, we adopt~\cite{wave0}, $|R_s(0)|^2=0.810\;{\rm GeV}^3$ and $|R'_p(0)|^2=0.075 \;{\rm GeV}^5$.

\subsection{Basic results}

\begin{table}
\begin{tabular}{|c||c|c|}
\hline
 $\sqrt{s}$ & ~~10.6 (GeV)~~ & ~~91.1876 (GeV)~~ \\
\hline
\hline
$\sigma_{e^+e^-\to \emph{J}/\psi \eta_{c}} $ & 5.967 & $3.5\times10^{-7}$\\
\hline
$\sigma_{e^+e^-\to \eta_{c} h_{c}} $ & 0.763 & $3.0\times10^{-6}$\\
\hline
$\sigma_{e^+e^-\to \emph{J}/\psi \chi_{c0}} $ & 7.011 & $2.0\times10^{-6}$\\
\hline
$\sigma_{e^+e^-\to \emph{J}/\psi \chi_{c1}} $ & 1.181 & $8.7\times10^{-8}$\\
\hline
$\sigma_{e^+e^-\to \emph{J}/\psi \chi_{c2}} $ & 1.703 & $3.5\times10^{-6}$\\
\hline
$\sigma_{e^+e^-\to \eta_{c} \gamma} $ & 67.90 & $1.3\times10^{-2}$ \\
\hline
$\sigma_{e^+e^-\to \chi_{c0} \gamma} $ & 1.855 & $5.5\times10^{-4}$ \\
\hline
$\sigma_{e^+e^-\to \chi_{c1} \gamma} $ & 20.69 & $3.3\times10^{-3}$ \\
\hline
$\sigma_{e^+e^-\to \chi_{c2} \gamma} $ & 8.138 & $1.1\times10^{-3}$ \\
\hline
\end{tabular}
\caption{Total cross sections (in unit: fb) for the charmonium production in $e^+e^-$ annihilation via a virtual photon at $\sqrt{s}=10.6$ GeV and $Z^0$-peak, respectively.}\label{tab1}
\end{table}

\begin{table}
\begin{tabular}{|c||c|c|}
\hline
 $\sqrt{s}$ & ~~10.6 (GeV)~~ & ~~91.1876 (GeV)~~ \\
\hline
\hline
$\sigma_{e^+e^-\to \emph{J}/\psi \eta_{c}} $ & $4.6\times10^{-5}$ & $2.0\times10^{-5}$ \\
\hline
$\sigma_{e^+e^-\to \eta_{c} h_{c}} $ & $8.0\times10^{-6}$ & $2.8\times10^{-4}$\\
\hline
$\sigma_{e^+e^-\to \emph{J}/\psi \chi_{c0}} $ & $5.4\times10^{-5}$ & $1.1\times10^{-4}$\\
\hline
$\sigma_{e^+e^-\to \emph{J}/\psi \chi_{c1}} $ & $9.0\times10^{-6}$ & $5.0\times10^{-6}$\\
\hline
$\sigma_{e^+e^-\to \emph{J}/\psi \chi_{c2}} $ & $1.3\times10^{-5}$ & $1.9\times10^{-4}$\\
\hline
$\sigma_{e^+e^-\to \emph{J}/\psi \gamma} $ & $4.0\times10^{-3}$ & 5.030\\
\hline
$\sigma_{e^+e^-\to \eta_{c} \gamma} $ & $5.2\times10^{-4}$ & 0.739 \\
\hline
$\sigma_{e^+e^-\to  h_{c} \gamma} $ & $4.6\times10^{-4}$ & 0.621 \\
\hline
$\sigma_{e^+e^-\to \chi_{c0} \gamma} $ & $1.4\times10^{-5}$ & 0.030 \\
\hline
$\sigma_{e^+e^-\to \chi_{c1} \gamma} $ & $1.6\times10^{-4}$ & 0.183 \\
\hline
$\sigma_{e^+e^-\to \chi_{c2} \gamma} $ & $7.0\times10^{-5}$ & 0.061 \\
\hline
\end{tabular}
\caption{Total cross sections (in unit: fb) for the charmonium production in $e^+e^-$ annihilation via the $Z^0$ boson at $\sqrt{s}=10.6$ GeV and $Z^0$-peak, respectively.}
\label{tab2}
\end{table}

\begin{table}
\begin{tabular}{|c||c|c|}
\hline
 ~~  ~~ & ~$\sqrt{s}=10.6$ (GeV)~ & ~$\sqrt{s}=91.1876$ (GeV)~ \\
\hline
\hline
$R_{e^+e^-\to \eta_{c} \gamma} $ & $0.04\%$ & $ 2.0\%$ \\
\hline
$R_{e^+e^-\to \emph{J}/\psi \gamma} $ & 0 & 0 \\
\hline
$R_{e^+e^-\to h_c \gamma} $ & 0 & 0 \\
\hline
$R_{e^+e^-\to \chi_{c0} \gamma} $ & $0.07\%$ & $3.4\%$\\
\hline
$R_{e^+e^-\to \chi_{c1} \gamma} $ & $0.07\%$ & $3.4\%$\\
\hline
$R_{e^+e^-\to \chi_{c2} \gamma} $ & $0.07\%$ & $2.8\%$\\
\hline
\end{tabular}
\caption{The ratio $R_X$ defined in Eq.(\ref{ratio}) for the single charmonium production in $e^+e^-$ annihilation at $\sqrt{s}=10.6$ GeV and $Z^0$-peak, respectively.}
\label{inter}
\end{table}

We put the total cross sections for the two charmonium production processes $e^+e^- \to H_{1}(c\bar{c})+H_{2}(c\bar{c})$ and $e^+e^- \to H(c\bar{c})+\gamma$ via the virtual photon or $Z^0$ boson in Tables \ref{tab1} and \ref{tab2}, respectively. To be useful reference, we take two collision energies to do our discussion. Here $\sqrt{s}=10.6$ GeV corresponds to the $B$ factory as BABAR or Belle, and $\sqrt{s}=m_Z=91.1876$ GeV corresponds to the so-called super $Z$ factory. In these two tables, the contribution from the interference terms for the channel via virtual photon and the channel via $Z^0$ boson has not been included. To show the relative importance of the interference cross sections, we define a ratio $R_{X}$
\begin{equation}\label{ratio}
R_{X}=\left.\frac{|\sigma_{\rm int}|}{\sigma_{\rm sep}}\right|_{X} ,
\end{equation}
where $X$ stands for the specific exclusive charmonium production channels, $\sigma_{\rm sep}$ is the cross section for the direct sum of the channel via virtual photon and $Z^0$ boson without interference contributions, $\sigma_{\rm int}$ is the interference cross section between the channel via virtual photon and the channel via $Z^0$ boson. As an explicit example, we put the values of $R_{X}$ for the single charmonium production in Table \ref{inter}. It shows that the interference terms only lead to small contributions. The cases for the double charmonium production are similar with even smaller values for $R_{X}$. So, in the present subsection, we will not consider the interference terms to provide a clear comparison of how the different channels will affect the cross-section at the $B$ factory and the super $Z$ factory, respectively.

Our present leading-order results for the total cross sections at the $B$ factories are consistent with those of Refs.\cite{plB557(2003)45-54,higher-charm,prdDL}, which is much smaller than the BABAR or Belle data. As has been stated in the Introduction, many suggestions, either the higher $\alpha_s$ order pQCD corrections or the relativistic corrections or the nonperturbative corrections or a combination of all these corrections, have been tried in the literature to explain the puzzle~\cite{prd67054007,prd77014002,sum rule,Braguta2008,plb57039,NLO corrections, relativistic, BLC2008, Wang20081,Wang20082,wang2012,wang2013}. Thus, a more accurate measurement or new measurements at another platform can be helpful for clarifying the puzzle.

At the $B$ factory, the charmonium production is dominated by the channels via the virtual photon, and the corresponding channels via the $Z^0$ boson are suppressed by about $10^4-10^7$ orders which can be safely neglected. On the other hand, at the super $Z$ factory, for the double charmonium production, both types of channels either via the virtual photon or via the $Z^0$ boson are around the same order and should be taken into consideration simultaneously, and for the single charmonium production, one only needs to deal with the channels via the $Z^0$ boson, since their cross sections are larger than those channels via the virtual photon by about three orders.

\begin{figure*}
\includegraphics[width=0.45\textwidth]{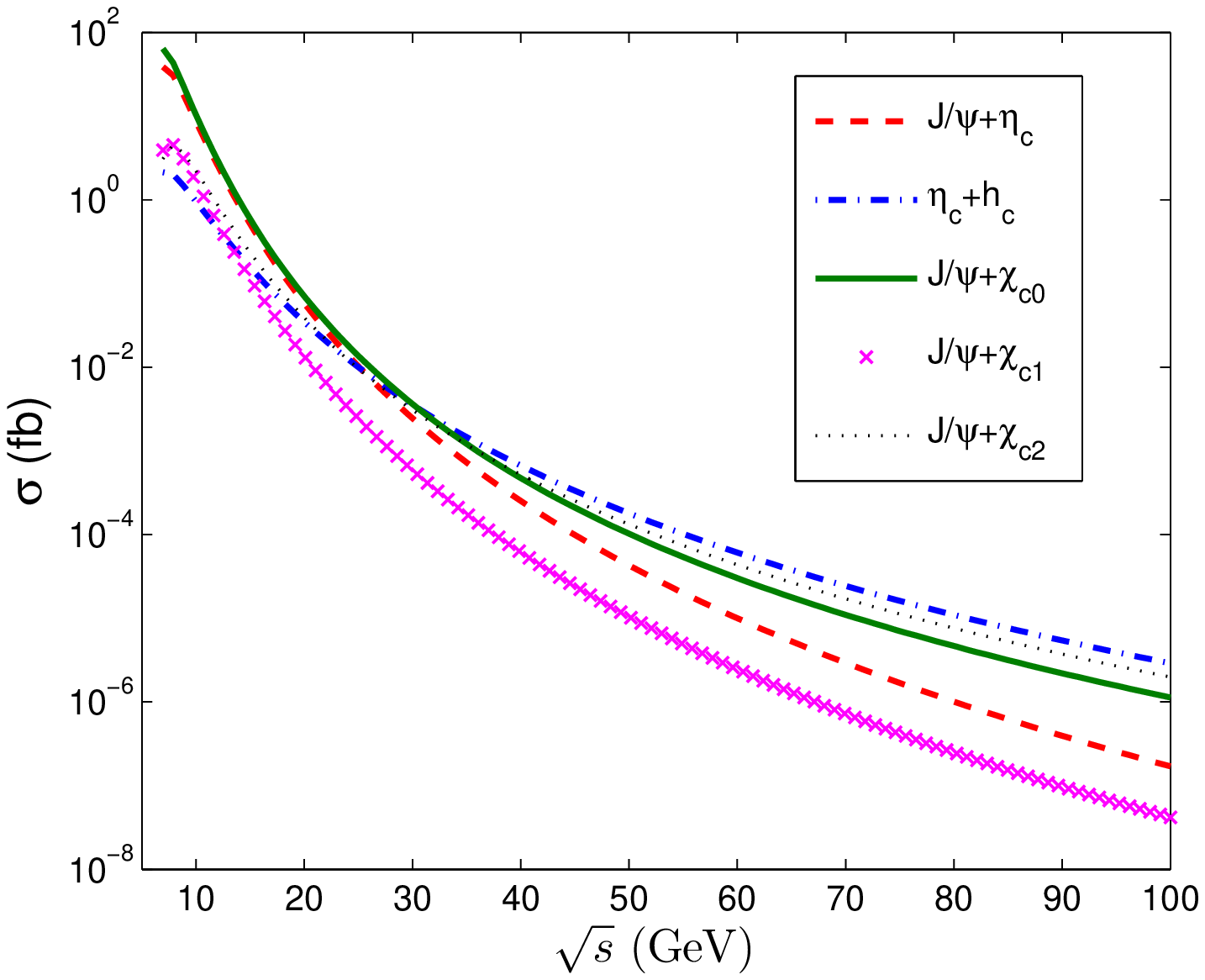}
\includegraphics[width=0.45\textwidth]{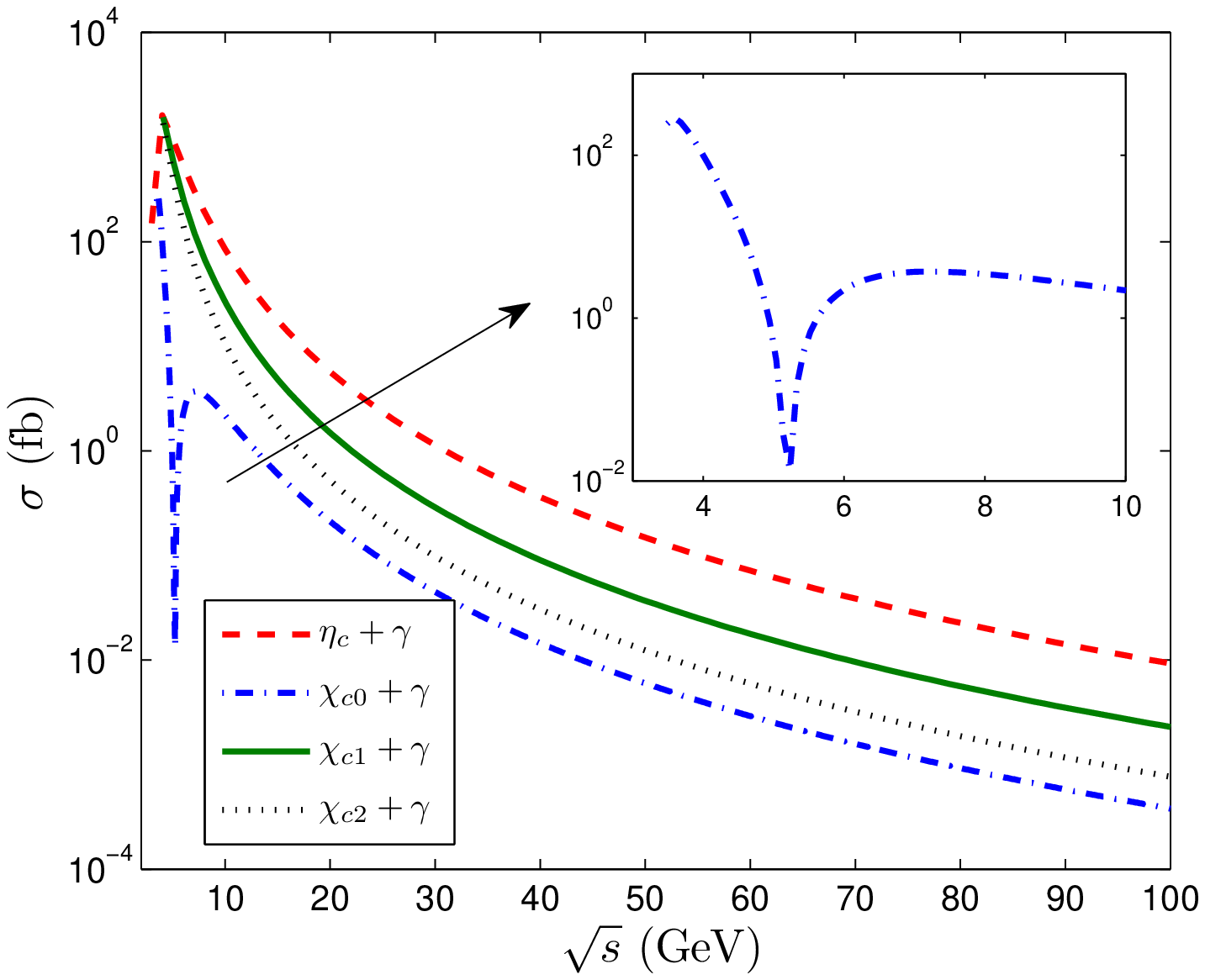}
\caption{Total cross sections versus the $e^{+}e^{-}$ collision energy $\sqrt{s}$ for the two channels via the virtual photon, $e^+e^-\to \gamma^* \to H_{1}(c\bar{c})+H_{2}(c\bar{c})$ and $e^+e^-\to \gamma^* \to H(c\bar{c})+\gamma$.}\label{tot1}
\end{figure*}

\begin{figure*}
\includegraphics[width=0.45\textwidth]{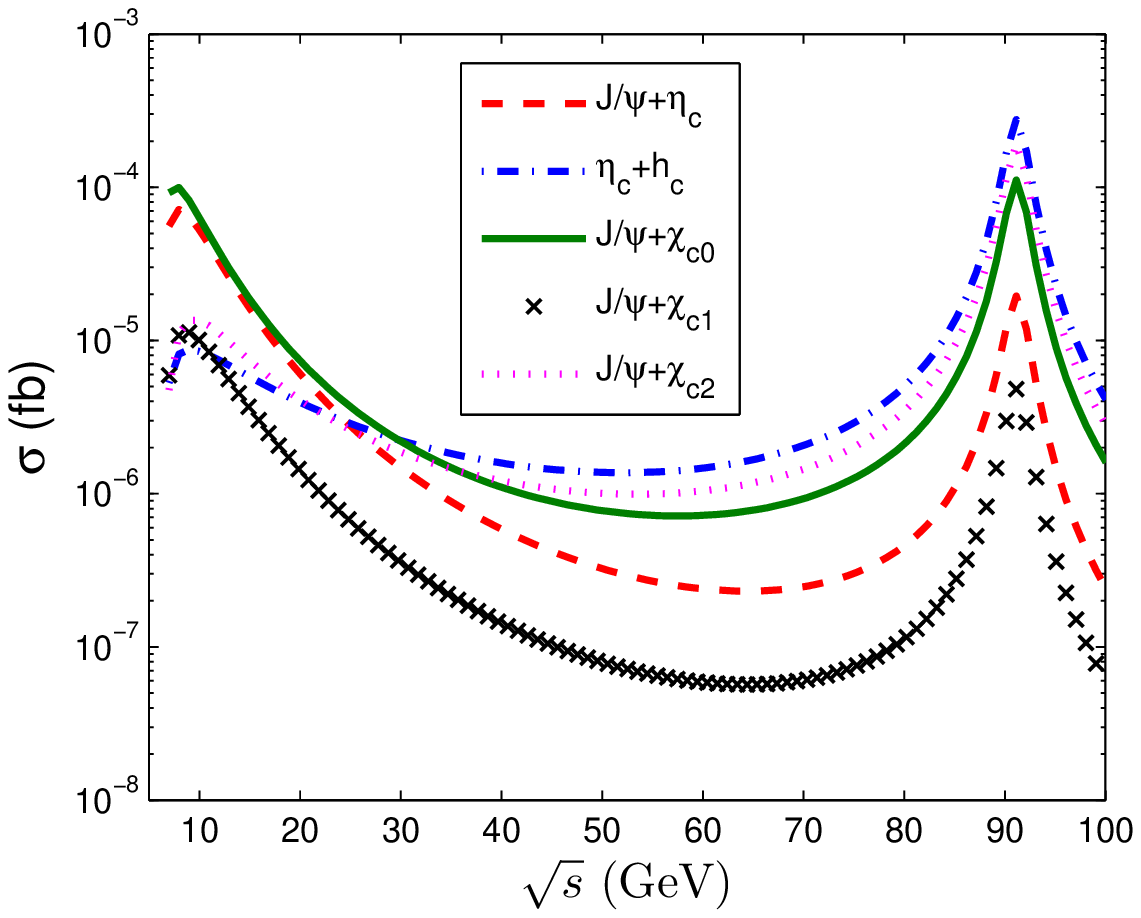}
\includegraphics[width=0.45\textwidth]{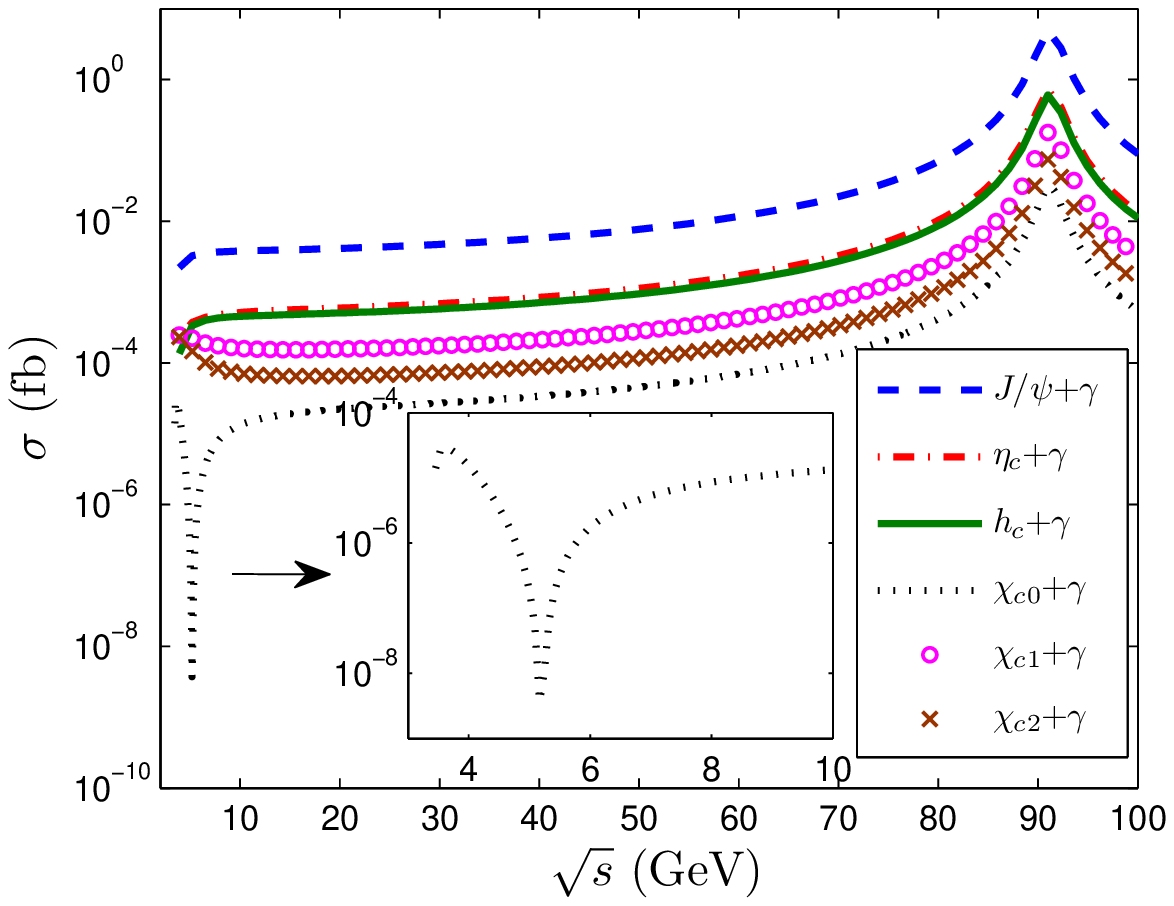}
\caption{Total cross sections versus the $e^{+}e^{-}$ collision energy $\sqrt{s}$ for the two channels via the $Z^0$ boson, $e^+e^-\to Z^0 \to H_{1}(c\bar{c})+H_{2}(c\bar{c})$ and $e^+e^-\to Z^0 \to H(c\bar{c})+\gamma$.}\label{tot2}
\end{figure*}

\begin{table}[h]
\begin{tabular}{|c||c|c|}
\hline
 $\sqrt{s}$ & ~~10.6 (GeV)~~ & ~~91.1876 (GeV)~~ \\
\hline
\hline
$\sigma_{e^+e^-\to \eta_{c} \chi_{c0}} $ & $5.7\times10^{-6}$ & $4.2\times10^{-4}$\\
\hline
$\sigma_{e^+e^-\to \eta_{c} \chi_{c1}} $ & $9.5\times10^{-5}$ & $4.9\times10^{-5}$\\
\hline
$\sigma_{e^+e^-\to \eta_{c} \chi_{c2}} $ & $6.7\times10^{-5}$ & $8.6\times10^{-4}$\\
\hline
$\sigma_{e^+e^-\to \emph{J}/\psi {J}/\psi} $ & $5.5\times10^{-5}$ & $4.4\times10^{-5}$\\
\hline
$\sigma_{e^+e^-\to \emph{J}/\psi h_c} $ & $1.2\times10^{-4}$ & $1.3\times10^{-3}$ \\
\hline
$\sigma_{e^+e^-\to  h_{c} \chi_{c0}} $ & $4.6\times10^{-6}$ & $2.1\times10^{-6}$ \\
\hline
$\sigma_{e^+e^-\to h_{c} \chi_{c1}} $ & $2.3\times10^{-6}$ & $4.3\times10^{-5}$ \\
\hline
$\sigma_{e^+e^-\to h_{c} \chi_{c2}} $ & $1.3\times10^{-6}$ & $1.4\times10^{-6}$ \\
\hline
$\sigma_{e^+e^-\to \chi_{c0} \chi_{c1}} $ & $1.3\times10^{-6}$ & $2.5\times10^{-5}$ \\
\hline
$\sigma_{e^+e^-\to \chi_{c0} \chi_{c2}} $ & $9.8\times10^{-8}$ & $2.2\times10^{-5}$ \\
\hline
$\sigma_{e^+e^-\to \chi_{c1} \chi_{c1}} $ & $5.9\times10^{-7}$ & $6.7\times10^{-7}$ \\
\hline
$\sigma_{e^+e^-\to \chi_{c1} \chi_{c2}} $ & $1.8\times10^{-7}$ & $5.3\times10^{-5}$ \\
\hline
$\sigma_{e^+e^-\to \chi_{c2} \chi_{c2}} $ & $2.2\times10^{-7}$ & $1.4\times10^{-5}$ \\
\hline
\end{tabular}
\caption{In additional to Table~\ref{tab2}, other non-zero cross sections (fb) for the double charmonium production via the $Z^0$-boson with $\sqrt{s}=m_Z$.} \label{tabnew}
\end{table}

We draw two figures, i.e. Fig.(\ref{tot1}) and Fig.(\ref{tot2}), to show the relative importance of those channels and to show how their total cross sections change with the $e^+ e^-$ collision energy. As for the channels $e^+e^-\to \gamma^* \to H_{1}(c\bar{c})+H_{2}(c\bar{c})$ and $e^+e^-\to \gamma^* \to H(c\bar{c})+\gamma$, their cross sections drop down logarithmically with the increment of the collision energy $\sqrt{s}$. As for the channel $e^+e^-\to Z^0 \to H_{1}(c\bar{c})+H_{2}(c\bar{c})$, it drops down logarithmically but with a much slower trends; while the production cross section of the channel $e^+e^-\to Z^0 \to H(c\bar{c})+\gamma$ slightly increases with the increment of $\sqrt{s}$. Both channels via $Z^0$ boson have a peak value at $\sqrt{s}=m_{Z}$ due to the $Z^0$-boson resonance effect. Thus, one may expect that those channels via $Z^0$ boson can provide sizable contributions at the super $Z$ factory.

More over, in Tables \ref{tab1} and \ref{tab2}, we have only shown those channels in which both types of production mechanisms via the virtual photon and the $Z^0$ boson have non-zero contributions. For the channels via the $Z^0$ boson, the final state combinations are much more involved, i.e. there are totally $21$ final-state combinations for $e^+e^-\to Z^0 \to H_{1}(c\bar{c})+H_{2}(c\bar{c})$, such as $^3S_1$+$^3S_1$ and etc.. In addition to those in Tables \ref{tab1} and \ref{tab2} , there are $16$ extra double charmonium combinations, which can not be happened due to the parity conservation in the production channels via the virtual photon propagator. For convenience, we put other non-zero cross sections for the double charmonium production via the $Z^0$ boson in Table \ref{tabnew}.

\begin{figure*}
\includegraphics[width=0.45\textwidth]{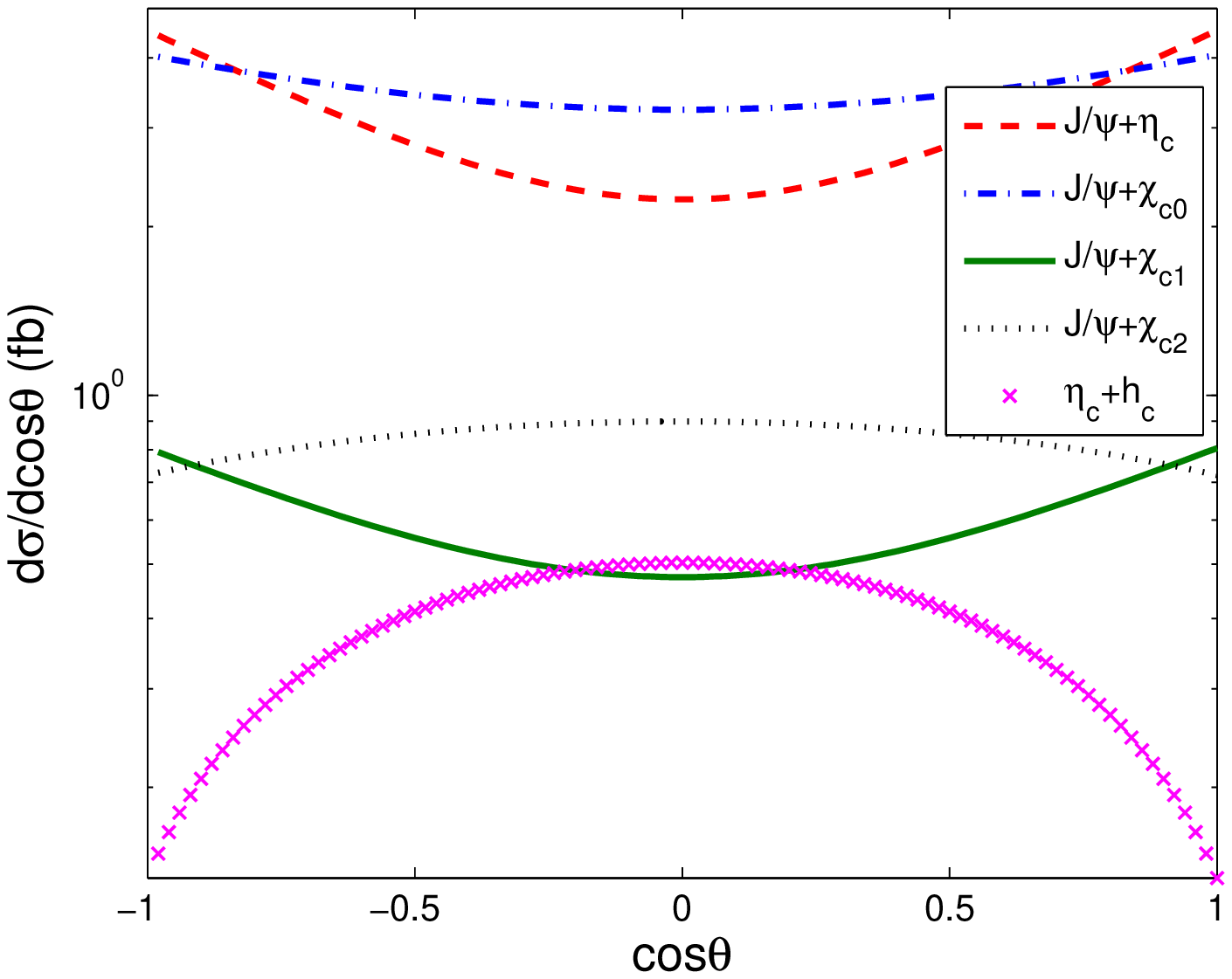}
\includegraphics[width=0.45\textwidth]{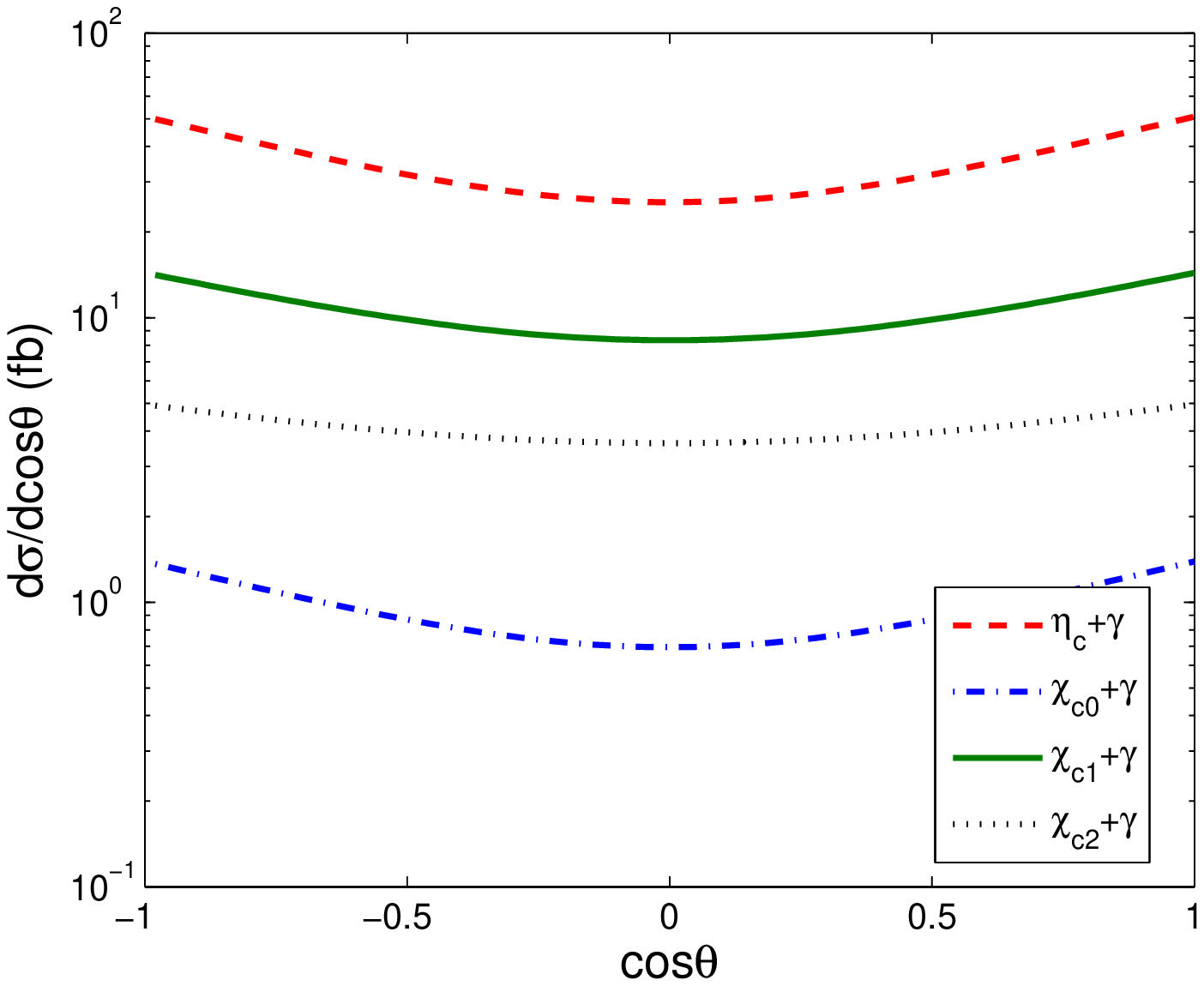}
\caption{Differential cross sections for the two channels via the virtual photon, $e^+e^-\to \gamma^* \to H_{1}(c\bar{c})+H_{2}(c\bar{c})$ and $e^+e^-\to \gamma^* \to H(c\bar{c})+\gamma$ at $\sqrt{s}=10.6$ GeV.}\label{diff1}
\end{figure*}

\begin{figure*}
\includegraphics[width=0.45\textwidth]{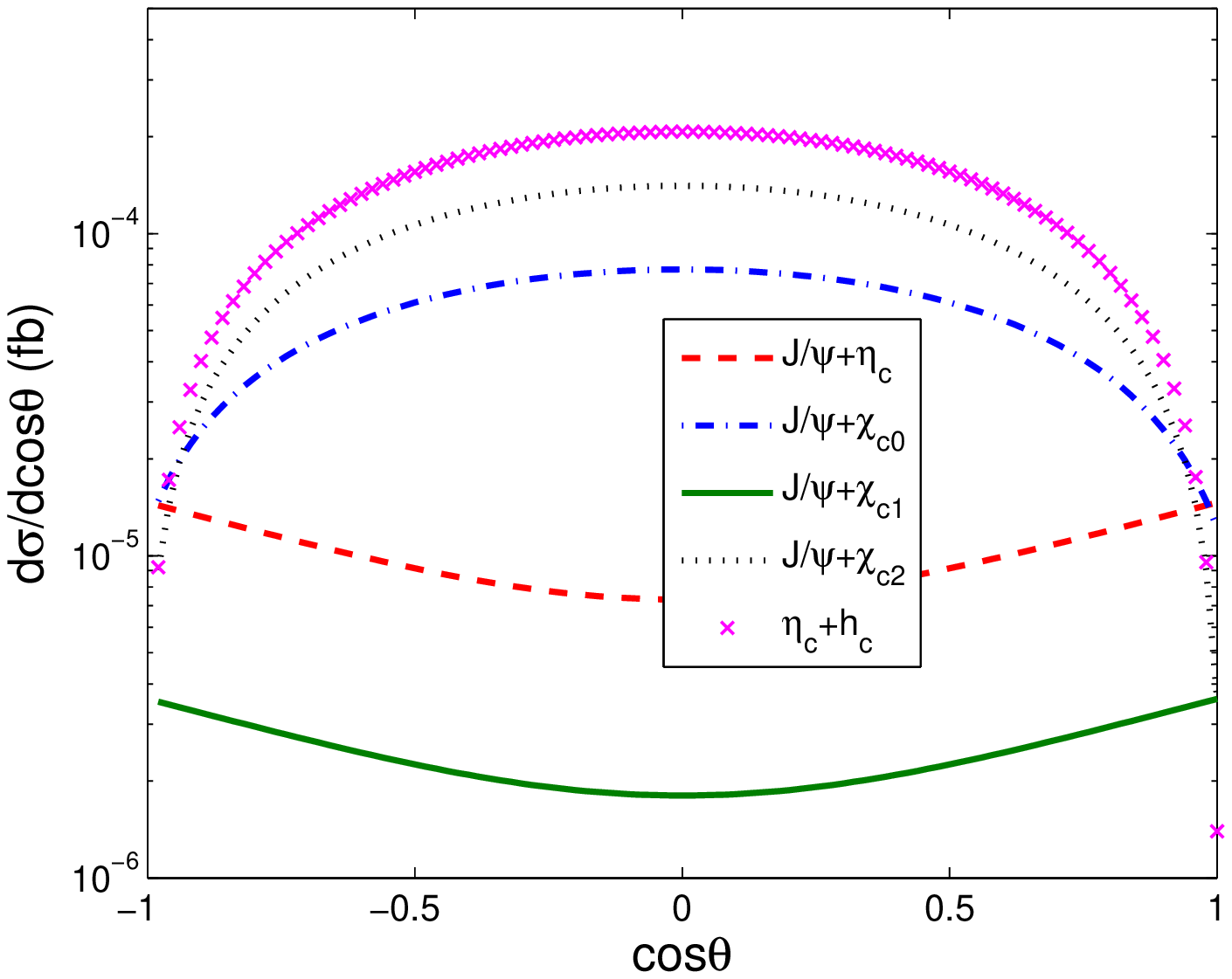}
\includegraphics[width=0.45\textwidth]{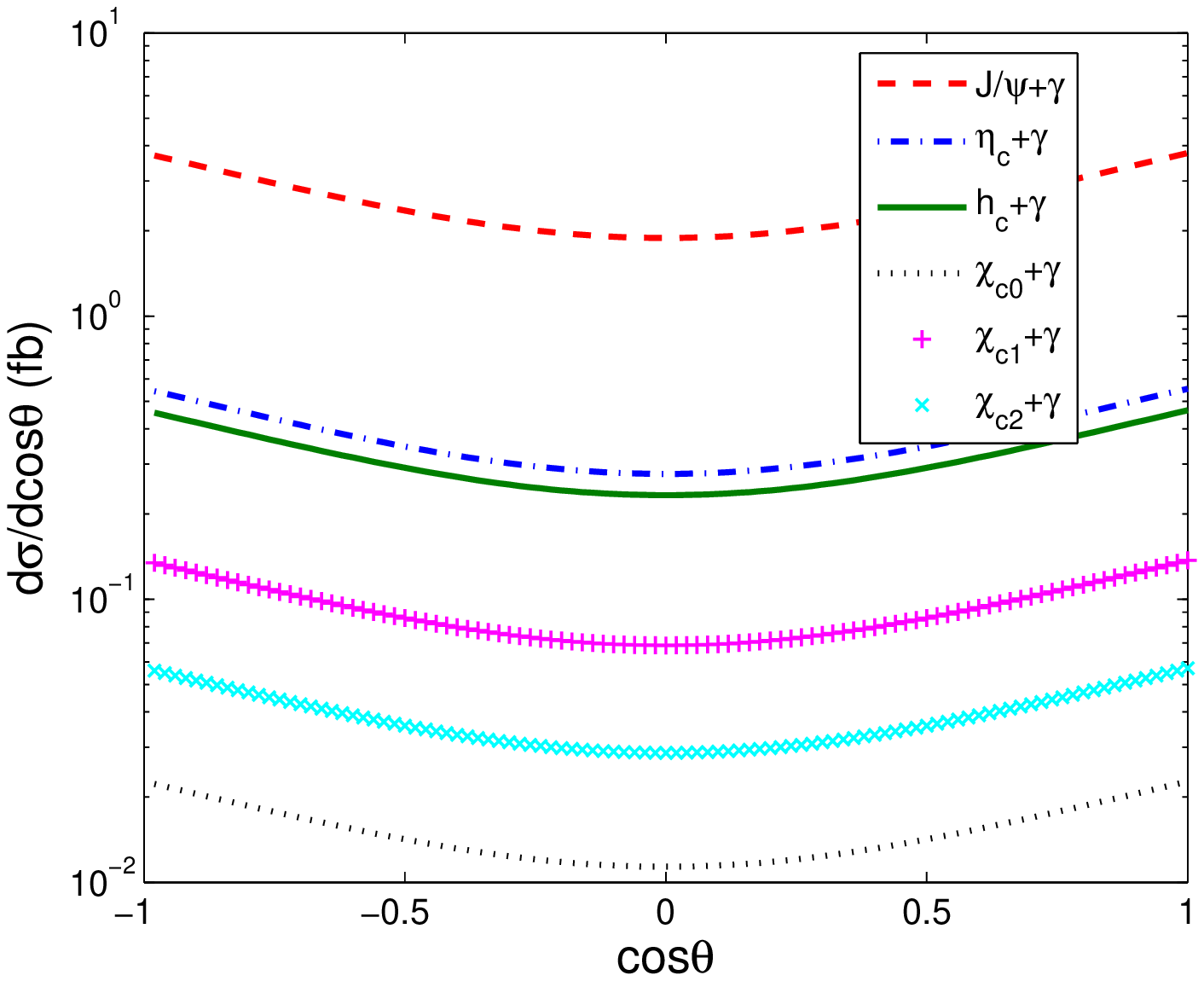}
\caption{Differential cross sections for the two channels via the $Z^0$ boson, $e^+e^-\to Z^0 \to H_{1}(c\bar{c})+H_{2}(c\bar{c})$ and $e^+e^-\to Z^0 \to H(c\bar{c})+\gamma$ at $\sqrt{s}=91.1876$ GeV.}\label{diff2}
\end{figure*}

We draw Fig.(\ref{diff1}) and Fig.(\ref{diff2}) to show the differential distributions for those processes versus $\cos\theta$, where $\theta$ is the angle between the three-vectors of the final charmonium ($\vec{q}_1$) and the electron $\vec{p}_1$. Those distributions show light concave behaviors in the whole kinematic region, the largest differential cross-section is at $\theta=0^{\circ}$ or $180^{\circ}$.

\subsection{Events to be generated at the super $Z$ factory}

According to the total cross sections, the super $Z$ factory running with high luminosity ${\cal L}\simeq 10^{36}$cm$^{-2}$s$^{-1}$ can provide another potential platform to study heavy quarkonium properties. For example, from the single charmonium production process $e^+ e^-\to \gamma^*/Z^0 \to H(c\bar{c})+\gamma$, we shall have $5.0\times10^4$ $J/\psi$, $7.5\times10^3$ $\eta_c$, $6.2\times10^3$ $h_{c}$, $3.1\times10^2$ $\chi_{c0}$, $2.2\times10^3$ $\chi_{c1}$, and $7.7\times10^2$ $\chi_{c2}$ events by one operation year. Totally, we shall have only $38.0$ double charmonium events generated through $e^+ e^-\to \gamma^*/Z^0 \to H_1(c\bar{c})+H_2(c\bar{c})$ by one operation year. In the estimation the channels via virtual photon and $Z^0$ boson together with their interference terms have been taken into consideration. This shows that at the super $Z$ factory, it is much more better to study the single charmonium processes other than the double charmonium processes.

\begin{figure*}
\includegraphics[width=0.44\textwidth]{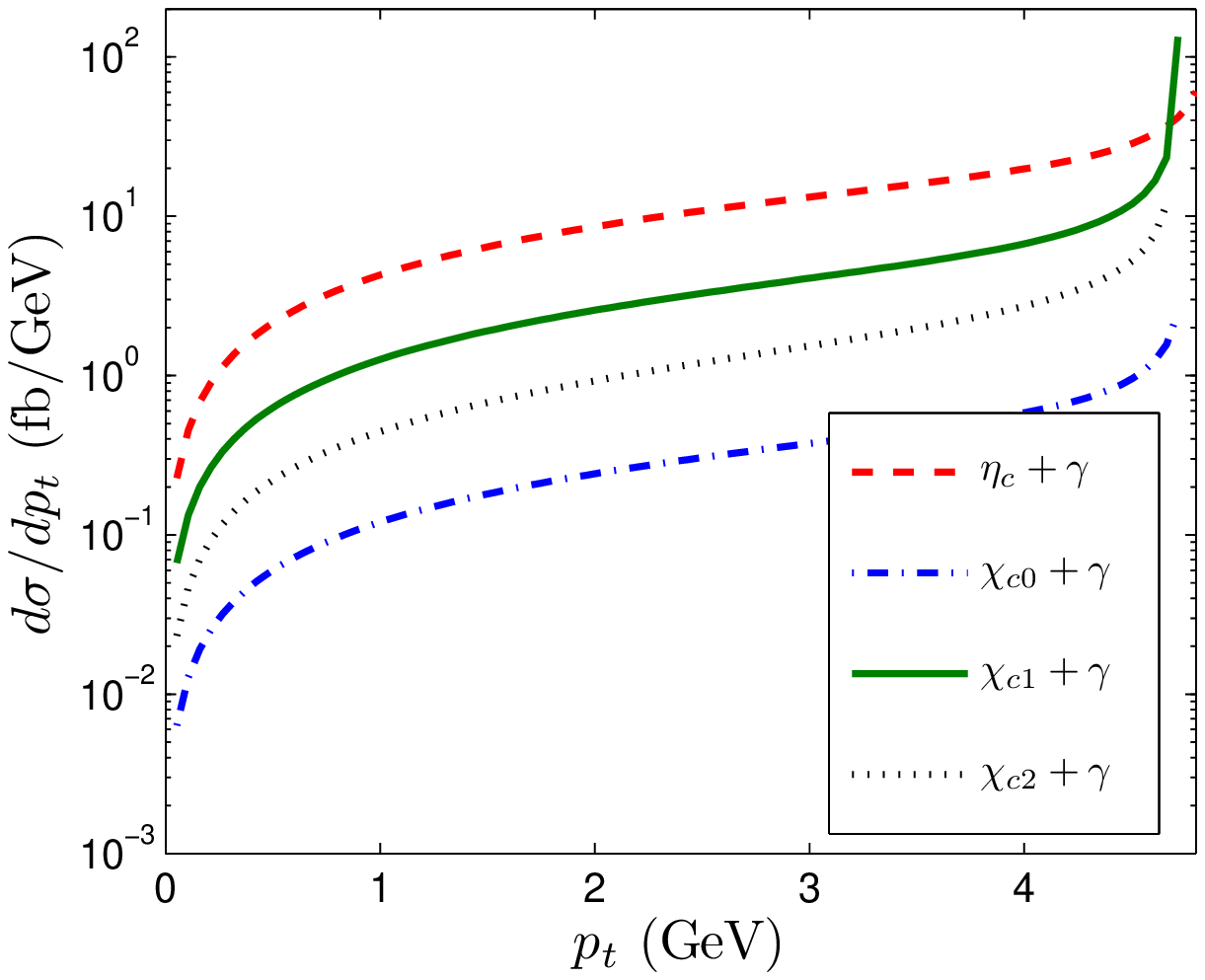}
\includegraphics[width=0.49\textwidth]{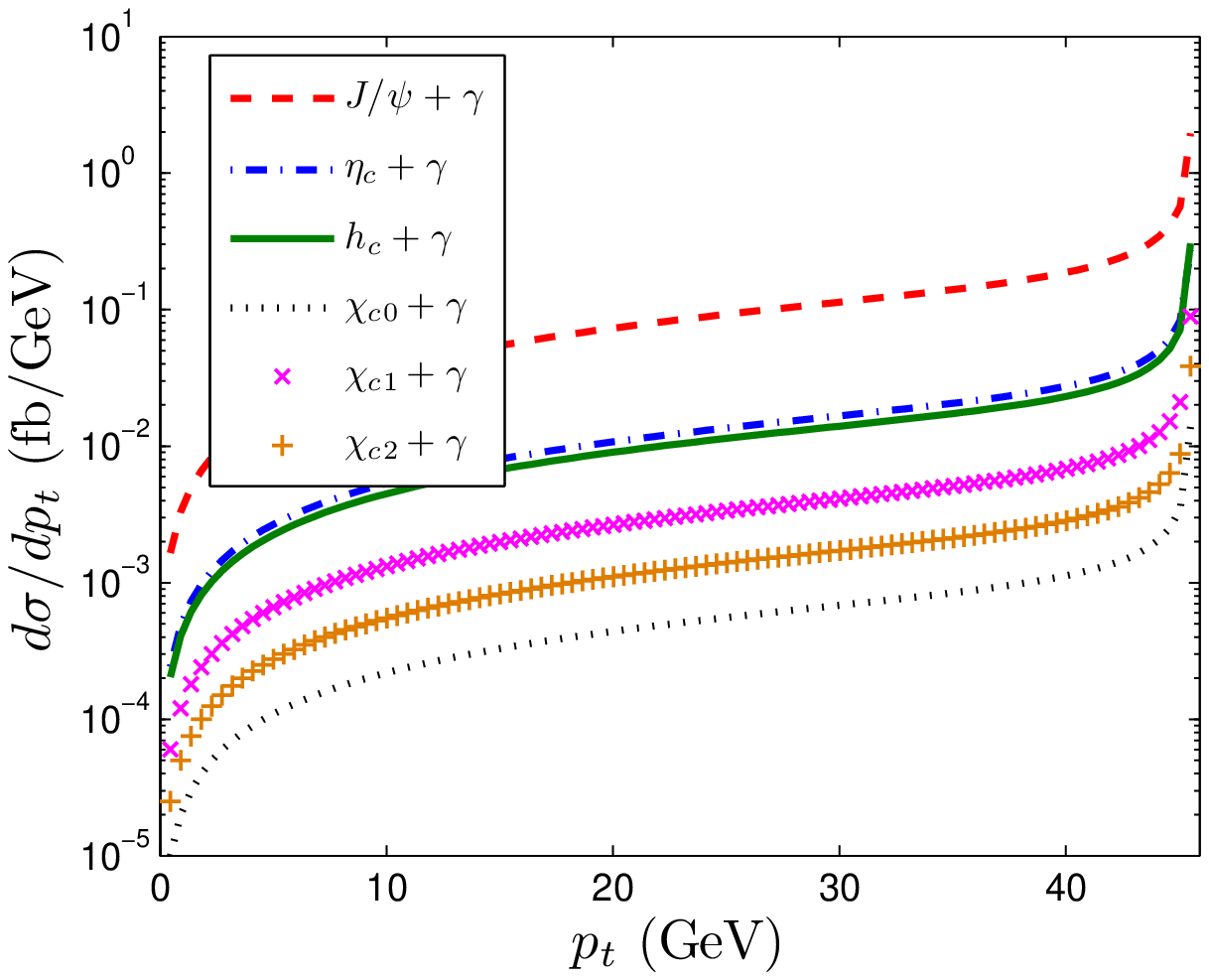}
\caption{Differential cross sections for $e^+e^-\to \gamma^*/Z^0 \to H(c\bar{c})+\gamma$ as functions of transverse momentum at $\sqrt{s}=10.6$ GeV and 91.1876 GeV, respectively.}\label{diff3}
\end{figure*}

Such number of $e^+(p_2) e^-(p_1) \to \gamma^*/Z^0 \to H(c\bar{c})(q_1)+\gamma(q_2)$ events make us possible to know more properties of the charmonium, such as its $p_t$ distributions. We can obtain the $p_t$ distributions from above $\cos\theta$-distributions through proper transformation. That is, if setting
\begin{equation}
\frac{d\sigma}{d\cos\theta}=f(\cos\theta),
\end{equation}
then, we obtain
\begin{eqnarray}
\frac{d\sigma}{dp_t} &=& \left(\frac{d\cos\theta}{dp_t}\right) \frac{d\sigma}{d\cos\theta} \nonumber\\
&=&  \left(\frac{- 2 p_t} {|\vec{q}_1| \sqrt{|\vec{q}_1|^2 - p_t^2}} \right) f\left(\frac{\sqrt{|\vec{q}_1|^2 - p_t^2}} {|\vec{q}_1|}\right)
\end{eqnarray}
where $|\vec{q}_{1}| = (s-M^2_{H(c\bar{c})})/ (2\sqrt{s})$. Here, we have implicitly used the symmetry $f(\cos\theta)=f(-\cos\theta)$ as indicated by Figs.(\ref{diff1},\ref{diff2}). We present the $p_t$ distributions for $e^+ e^-\to \gamma^*/Z^0 \to H(c\bar{c})+\gamma$ in Fig.(\ref{diff3}), where $p_t$ is the transverse momentum of the charmonium. We have largest $f(\cos\theta)$ when the charmonium and the photon running along the same direction or rightly back-to-back, however, because of the phase-space factor proportional to ${p_t}/ {\sqrt{|\vec{q}_1|^2 - p_t^2}}$, the $p_t$ distribution shall increase with the increment of $p_t$.

As a final remark, as shown by the right diagrams of Figs.(\ref{tot1},\ref{tot2}), there is a dip behavior at $\sqrt{s}\simeq 5.2$ GeV for the $\chi_{c0}$ production via the process $e^+e^-\to \gamma^*/Z^0 \to \chi_{c0}+\gamma$. This is caused by the overall suppression factor $(s-12 m_{c}^2)^2$, which does not appear in the other production channels and approaches zero when $\sqrt{s}\simeq 5.2$ GeV for $m_c=1.5$ GeV.

\subsection{A simple analysis of theoretical uncertainty caused by the charm quark mass}

\begin{table}
\begin{tabular}{|c||c|c|c|}
\hline
 $m_{c}$ & ~~$1.3$ (GeV)~~ & ~~$1.5$ (GeV)~~ & ~~$1.7$ (GeV)~~ \\
\hline
\hline
$\sigma_{\gamma^* \to \emph{J}/\psi \eta_{c}} $ & 6.656 & 5.957 & 5.158 \\
\hline
$\sigma_{\gamma^* \to \eta_{c} h_{c}} $ & 1.613 & 0.763 & 0.384 \\
\hline
$\sigma_{\gamma^* \to \emph{J}/\psi \chi_{c0}} $ & 10.04 & 7.011 & 5.059 \\
\hline
$\sigma_{\gamma^* \to \emph{J}/\psi \chi_{c1}} $ & 1.846 & 1.181 & 0.750 \\
\hline
$\sigma_{\gamma^* \to \emph{J}/\psi \chi_{c2}} $ & 3.068 & 1.703 & 0.950 \\
\hline
\hline
$\sigma_{\gamma^* \to \eta_{c} \gamma} $ & 78.38 & 67.93 & 59.93 \\
\hline
$\sigma_{\gamma^* \to \chi_{c0} \gamma} $ & 3.470 & 1.856 & 1.001 \\
\hline
$\sigma_{\gamma^* \to \chi_{c1} \gamma} $ & 31.06 & 20.70 & 14.65 \\
\hline
$\sigma_{\gamma^* \to \chi_{c2} \gamma} $ & 12.13 & 8.144 & 5.841 \\
\hline
\end{tabular}
\caption{Uncertainties caused by the $c$-quark mass for the total cross-sections (in unit: fb) of the charmonium production with the center-of-mass energy $\sqrt{s}=10.6$ GeV, in which the channels via virtual photon and $Z^0$ boson together with their interference terms have been taken into consideration. }
\label{tabrpb1}
\end{table}

\begin{table}
\begin{tabular}{|c||c|c|c|}
\hline
 $m_{c}$ & ~~$1.3$ (GeV)~~ & ~~$1.5$ (GeV)~~ & ~~$1.7$ (GeV)~~ \\
\hline
\hline
$\sigma_{Z^0\to \emph{J}/\psi \eta_{c}} $ & 2.0$\times 10^{-5}$ & 2.0$\times 10^{-5}$ & 2.0$\times 10^{-5}$\\
\hline
$\sigma_{Z^0\to \eta_{c} h_{c}} $ & 4.9$\times 10^{-4}$ & 2.8$\times 10^{-4}$ & 1.7$\times 10^{-4}$ \\
\hline
$\sigma_{Z^0\to \emph{J}/\psi \chi_{c0}} $ & 1.9$\times 10^{-4}$ & 1.1$\times 10^{-4}$ & 7.1$\times 10^{-5}$ \\
\hline
$\sigma_{Z^0\to \emph{J}/\psi \chi_{c1}} $ & 6.4$\times 10^{-6}$ & 4.8$\times 10^{-6}$ & 3.7$\times 10^{-6}$ \\
\hline
$\sigma_{Z^0\to \emph{J}/\psi \chi_{c2}} $ & 3.4$\times 10^{-4}$ & 1.9$\times 10^{-4}$ & 1.2$\times 10^{-4}$ \\
\hline
\hline
$\sigma_{Z^0\to \emph{J}/\psi \gamma} $ & 5.799 & 5.030 & 4.442 \\
\hline
$\sigma_{Z^0\to \eta_{c} \gamma} $ & 0.885 & 0.767 & 0.677 \\
\hline
$\sigma_{Z^0\to  h_{c} \gamma} $ & 0.953 & 0.621 & 0.427 \\
\hline
$\sigma_{Z^0\to \chi_{c0} \gamma} $ & 0.049 & 0.032 & 0.021 \\
\hline
$\sigma_{Z^0\to \chi_{c1} \gamma} $ & 0.296 & 0.193 & 0.132 \\
\hline
$\sigma_{Z^0\to \chi_{c2} \gamma} $ & 0.122 & 0.080 & 0.055 \\
\hline
\end{tabular}
\caption{Uncertainties caused by the $c$-quark mass for the total cross-sections (in unit: fb) of the charmonium production with the center-of-mass energy $\sqrt{s}=91.1876$ GeV, in which the channels via virtual photon and $Z^0$ boson together with their interference terms have been taken into consideration. }
\label{tabrpb2}
\end{table}

In this subsection, we do a simple analysis of the theoretical uncertainty caused by the charm quark mass.

We present the total cross sections with $m_c=1.50\pm0.20$ GeV in Tables \ref{tabrpb1} and \ref{tabrpb2}, in which the channels via virtual photon and $Z^0$ boson together with their interference terms have been taken into consideration. Total cross sections of those production channels are sensitive to the $c$-quark mass, which increase with the decrement of $m_c$. More explicitly, we present the uncertainties for the case of $\sqrt{s}=10.6$ GeV in the following,
\begin{eqnarray}
\sigma_{e^+e^-\to \emph{J}/\psi \eta_{c}} &=& 5.957^{+0.699}_{-0.799}\;{\rm fb},\\
\sigma_{e^+e^-\to \eta_{c} h_{c}}&=& 0.763^{+0.850}_{-0.379}\;{\rm fb},\\
\sigma_{e^+e^-\to \emph{J}/\psi \chi_{c0}}&=&7.011^{+3.029}_{-1.952}\;{\rm fb},\\
\sigma_{e^+e^-\to \emph{J}/\psi \chi_{c1}}&=& 1.181^{+0.665}_{-0.431}\;{\rm fb},\\
\sigma_{e^+e^-\to \emph{J}/\psi \chi_{c2}}&=&1.703^{+1.365}_{-0.753}\;{\rm fb},\\
\sigma_{e^+e^-\to \eta_{c} \gamma}&=&67.93^{+10.45}_{-8.00}\;{\rm fb},\\
\sigma_{e^+e^-\to \chi_{c0} \gamma}&=&1.855^{+1.614}_{-0.855}\;{\rm fb},\\
\sigma_{e^+e^-\to \chi_{c1} \gamma}&=&20.69^{+10.36}_{-6.050}\;{\rm fb},\\
\sigma_{e^+e^-\to \chi_{c2} \gamma}&=&8.138^{+3.986}_{-2.303}\;{\rm fb}.
\end{eqnarray}
Simultaneously, we present the uncertainties for the case of $\sqrt{s}=91.1876$ GeV in the following,
\begin{eqnarray}
\sigma_{e^+e^-\to \emph{J}/\psi \eta_{c}} &=& 2.0^{+0.002}_{-0.002}\times 10^{-5}\;{\rm fb},\\
\sigma_{e^+e^-\to \eta_{c} h_{c}}&=& 2.8^{+2.1}_{-1.1}\times 10^{-5}\;{\rm fb},\\
\sigma_{e^+e^-\to \emph{J}/\psi \chi_{c0}}&=&1.1^{+0.8}_{-0.4}\times 10^{-5}\;{\rm fb},\\
\sigma_{e^+e^-\to \emph{J}/\psi \chi_{c1}}&=& 4.8^{+1.6}_{-1.1}\times 10^{-5}\;{\rm fb},\\
\sigma_{e^+e^-\to \emph{J}/\psi \chi_{c2}}&=&1.9^{+1.5}_{-0.7}\times 10^{-5}\;{\rm fb},\\
\sigma_{e^+e^-\to \emph{J}/\psi \gamma}&=&5.030^{+0.769}_{-0.588}\;{\rm fb},\\
\sigma_{e^+e^-\to \eta_{c} \gamma}&=&0.767^{+0.118}_{-0.090}\;{\rm fb},\\
\sigma_{e^+e^-\to  h_{c} \gamma}&=&0.621^{+0.332}_{-0.194}\;{\rm fb},\\
\sigma_{e^+e^-\to \chi_{c0} \gamma}&=&0.032^{+0.017}_{-0.011}\;{\rm fb},\\
\sigma_{e^+e^-\to \chi_{c1} \gamma}&=&0.193^{+0.103}_{-0.061}\;{\rm fb},\\
\sigma_{e^+e^-\to \chi_{c2} \gamma}&=&0.080^{+0.042}_{-0.025}\;{\rm fb}.
\end{eqnarray}
It is noted that among the single charmonium production processes at the super $Z$ factory, the cross section $\sigma_{e^+e^-\to \emph{J}/\psi \gamma}$ is much larger than the production of other charmonium states. The production cross section for $\emph{J}/\psi \gamma$ is zero for the channel via a virtual photon due to Landau-Pomeranchuk-Yang theorem~\cite{yang}; while, for the channel via $Z^0$-boson, the pseudo-vector vertex $\gamma_\mu\gamma_5$ will lead to non-zero contributions, which furthermore do not have the suppression factor $(1-4\sin^2\theta)$ as the case of the vector vertex $\gamma_\mu$ part that dominantly determines other charmonium states' production.

\section{Summary}

In the present paper, we have studied the charmonium exclusive production through $e^+e^-$ annihilation at the collision energy that equals to the $B$-factories or the super $Z$-factory.

As shown by Tables \ref{tab1} and \ref{tab2}: 1) At the $B$ factory, the charmonium production is dominated by the channels via the virtual photon, since the total cross-sections for the single charmonium production channel are at the order of $10^{-6}-10^{-5}$ fb; 2) At the super $Z$ factory, the total cross-sections are dominated by the single charmonium production channel. For the double charmonium production, both types of channels either via the virtual photon or via the $Z^0$-boson propagator are around the same order, about $10^{-7}-10^{-4}$ fb, and for the single charmonium production, one may only need to deal with the channels via the $Z^0$ boson (the interference terms between the channels via virtual photon and the $Z^0$-boson propagator will bring about several percent corrections as shown by Table \ref{inter}.).

We have drawn Fig.(\ref{tot1}) and Fig.(\ref{tot2}) to show the relative importance of the production channels and to show how their total cross sections change with the $e^+ e^-$ collision energy. Because the channels $e^+e^-\to Z^0 \to H_{1}(c\bar{c})+H_{2}(c\bar{c})$ and $e^+e^-\to Z^0 \to H(c\bar{c})+\gamma$ have a peak value at $\sqrt{s}=m_{Z}$ due to the $Z^0$-boson resonance effect, one may expect that those channels via $Z^0$ boson can provide sizable contributions at the super $Z$ factory.

\begin{figure}[h]
\includegraphics[width=0.45\textwidth]{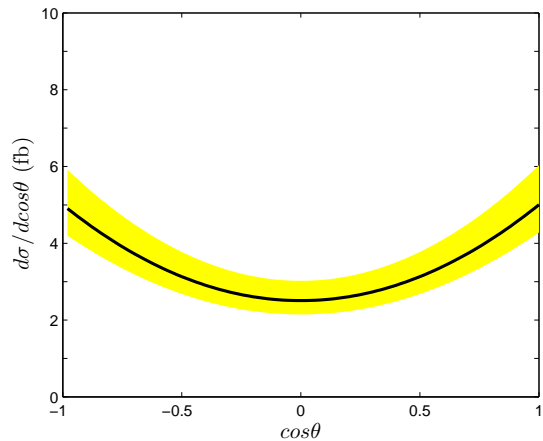}
\caption{Uncertainties of $d\sigma/d\cos\theta$ for $e^+ e^-\to H(c\bar{c}) + \gamma$ at the super $Z$ factory, where contributions from the color-singlet $S$-wave and $P$-wave states have been summed up. The upper edge of the band is for $m_c=1.30$ GeV, the lower edge of the band is for $m_c=1.70$ GeV, and the central line is for $m_c=1.50$ GeV.}\label{unc1}
\end{figure}

\begin{figure}[h]
\includegraphics[width=0.45\textwidth]{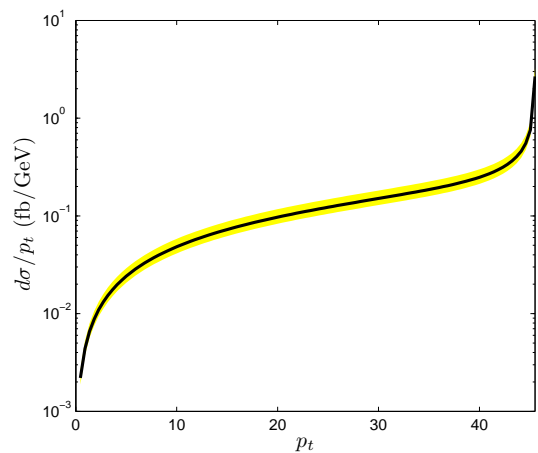}
\caption{Uncertainties of $d\sigma/dp_t$ for $e^+ e^-\to H(c\bar{c}) + \gamma$ at the super $Z$ factory, where contributions from the color-singlet $S$-wave and $P$-wave states have been summed up. The upper edge of the band is for $m_c=1.30$ GeV, the lower edge of the band is for $m_c=1.70$ GeV, and the central line is for $m_c=1.50$ GeV.}\label{unc2}
\end{figure}

It is shown that the super $Z$ factory running with high luminosity ${\cal L}\simeq 10^{36}$cm$^{-2}$s$^{-1}$ can provide another potential platform to study heavy quarkonium properties. As for the single charmonium production process $e^+ e^-\to \gamma^*/Z^0 \to H(c\bar{c})+\gamma$, by taking $m_c=1.50\pm0.20$ GeV, we shall have $(5.0^{+0.8}_{-0.6})\times10^4$ $J/\psi$, $(7.5^{+1.1}_{-0.9})\times10^3$ $\eta_c$, $(6.2^{+3.3}_{-1.9})\times10^3$ $h_{c}$, $(3.1^{+1.7}_{-0.9})\times10^2$ $\chi_{c0}$, $(2.2^{+1.0}_{-0.4})\times10^3$ $\chi_{c1}$, and $(7.7^{+4.1}_{-2.4})\times10^2$ $\chi_{c2}$ events by one operation year. Sizable events make us possible to know more properties of the channel, for examples,
\begin{itemize}
\item The symmetric charmonium angle distributions $d\sigma/d\cos\theta$ are shown in Figs.(\ref{diff1},\ref{diff2}). Those distributions show light concave behaviors in the whole kinematic region, the largest differential cross-section is at $\theta=0^{\circ}$ or $180^{\circ}$.

\item The charmonium transverse momentum distributions $d\sigma/dp_t$ are shown in Fig.(\ref{diff3}), in which it shows the distributions for the single charmonium production increase with the increment of $p_t$ due to the phase-space enhancement ${p_t}/ {(\sqrt{|\vec{q}_1|^2 - p_t^2})}$ at the large $p_t$ region.

\item To show how the single charmonium production cross-sections depend on the charm-quark masses, we present its differential cross-sections $d\sigma/d\cos\theta$ and $d\sigma/dp_t$ with $m_c=1.50\pm0.20$ GeV in Figs.(\ref{unc1},\ref{unc2}). In these two figures, the contributions from the color-singlet $S$-wave and $P$-wave states have been summed up. The higher excited charmonium states may decay to the ground color-singlet and spin-singlet $S$ wave state with $100\%$ efficiency via the electromagnetic or the hadronic interactions.

\end{itemize}

Our present calculation is at the leading-order level. It could be helpful to finish a next-to-leading order calculation on those processes at the super $Z$ factory, in which we have to deal with the extension of $\gamma_5$ to any dimensions and to show how to do the dimensional analysis within the improved trace technology and also to set the renormalization scale properly so as to eliminate the scale ambiguity at the known fixed perturbative order, which are in progress~\cite{chen}. \\

{\bf Acknowledgement:} This work was supported in part by the Fundamental Research Funds for the Central Universities under Grant No.CQDXWL-2012-Z002, by Natural Science Foundation of China under Grant No.11075225 and No.11275280, and by the Program for New Century Excellent Talents in University under Grant NO.NCET-10-0882.

\appendix

\begin{widetext}

\section{Independent Lorentz structures and their coefficients for $\bm{e^{+}+e^{-} \to Z^0 \to H_{1}(c\bar{c}) + H_{2}(c\bar{c})}$}

For useful reference, we present the independent Lorentz structures and their coefficients for the typical production processes mentioned in the body of the text. The program is available upon request. We use two extra symmetries to simplify the analytic expressions: 1) there is $u\leftrightarrow t$ symmetry; 2) there is $s\leftrightarrow -s$ symmetry. In addition, the properties of the $^3P_J$ states's polarization tensor $\varepsilon^{J}_{\alpha\beta}$ are useful, i.e. $\varepsilon^{0,2}_{\alpha\beta}$ is symmetric, $\varepsilon^{1}_{\alpha\beta}$ is anti-symmetric, and $\varepsilon^{1}_{\alpha\alpha}=\varepsilon^{2}_{\alpha\alpha}=0$.

In this section, we present the Lorentz structures $B_j$ and their nonzero coefficients $A^{n}_j$ for the process $e^+(p_{2})+e^-(p_{1})\to Z^{0}\to H_{1} (c\bar{c})(q_{1})+H_{1}(c\bar{c})(q_{2})$, where $n=(1,\cdots,4)$, $j=(1,\cdots,\eta)$ with $\eta$ the maximum number of basic Lorentz structures, the quarkonium state can be taken as $|(c\bar{c})_{\bf 1}[^1S_0]\rangle$, $|(c\bar{c})_{\bf 1}[^3S_1]\rangle$, $|(c\bar{c})_{\bf 1}[^1P_1]\rangle$, and $|(c\bar{c})_{\bf 1}[^3P_J]\rangle(J=1,2,3)$, respectively. Since $m_{e}\ll\sqrt{s}$, we have found that the electron mass effect is small, so we shall neglect it in the following analytic expressions. Especially, under such approximation, we shall always have $A^{1,2}_j \equiv 0$.

For convenience, we define some notations that are frequently emerged in the non-zero coefficients,
\begin{eqnarray}
&& s_{w}={\sin ^2}{\theta _w},\;\;
L_{1}=L_{2}=\frac{{\cal C}}{\sqrt{2s}},\;\;
\kappa = \frac{2}{\sqrt{16 m_{c}^4 s-s t u}},\nonumber\\
&& d_{1}= \frac{1}{2 s^2 (1-s_{w}) \sqrt{m_{z}^4+\Gamma_z^2 m_{z}^2-2 m_{z}^2 s+s^2}},\;\;
d_{2}=\frac{2d_{1}}{s},\;\;
d_{3}= \frac{4d_{1}}{s}.\nonumber
\end{eqnarray}
where, $s=(p_1 +p_2)^2$, $t=(p_1 -q_1)^2$, and $u=(p_1-q_2)^2$ are Mandelstam variables. The symbol $\varepsilon(\alpha, \beta, p, q)$ used in the Lorentz structures is defined as $\varepsilon(\alpha, \beta, p, q)=\varepsilon^{\alpha \beta \mu \nu} p_{\mu} q_{\nu}$, where $\varepsilon^{\alpha \beta \mu \nu}$ is the antisymmetric tensor, $\alpha$ and $\beta$ are lorentz indices, $p$ and $q$ are corresponding momenta.

\subsection{$e^+(p_2) e^-(p_1) \to Z^0 \to J/\psi(q_1) + \eta_{c}(q_2)$}

There are four independent Lorentz structures for the process, which are
\begin{eqnarray}
B_{1}&=&\frac{i}{m_{c}^3} \varepsilon(\alpha, p_1, p_2, q_1),\;\;
B_{2}=\frac{p_{1}^{\alpha }}{m_c},\;\;\nonumber
B_{3}=\frac{p_{2}^{\alpha }}{m_c},\;\;
B_{4}=\frac{q_{1}^{\alpha }}{m_c},\;\;\nonumber
\end{eqnarray}
whose non-zero coefficients $A^{3,4}_j$ with $j=(1,\cdots,4)$ are
\begin{eqnarray}
A^{3}_1 &=& \frac{4 i m_c^3 \kappa}{3 L_1} (8 s_{w}-3) (t-u),\\
A^{3}_2 &=& -A^{3}_{3}|_{t\leftrightarrow u} = -\frac{2 i m_c \kappa s}{3 L_1} (4 s_{w}-1) (8 s_{w}-3) (4 m_c^2 + t),\\
A^{3}_4 &=& \frac{2 i m_c \kappa s}{3 L_1} (4 s_{w}-1) (8 s_{w}-3) (t-u),\\
A^{4}_1 &=& \frac{2m_c^2}{s} A^{3}_4, \;\;
A^{4}_2 = -A^{4}_{3}|_{t\leftrightarrow u} = (4 s_{w}-1)A^{3}_4,\;\;
A^{4}_4 = \frac{s}{2m_c^2} A^{3}_1.
\end{eqnarray}

\subsection{$ e^+(p_2) e^-(p_1) \to Z^0 \to \eta_{c}(q_1) + h_{c}(q_2)$}

There are four independent Lorentz structures for this process, which are
\begin{eqnarray}
B_{1}&=&\frac{i}{m_{c}^3} \varepsilon(\alpha, p_1, p_2, q_1),\;\;
B_{2}=\frac{p_{1}^{\alpha }}{m_c},\;\;\nonumber
B_{3}=\frac{p_{2}^{\alpha }}{m_c},\;\;
B_{4}=\frac{q_{1}^{\alpha }}{m_c},\;\;\nonumber
\end{eqnarray}
whose non-zero coefficients $A^{3,4}_j$ with $j=(1,\cdots,4)$ are
\begin{eqnarray}
A^{3}_1 &=& \frac{i d_1 m_{c}^2 \kappa}{3 L_1} (16 m_{c}^2 (4 s_{w}-1) (8 s_{w}-3)+s (8 (5-8 s_{w}) s_{w}-6)-3 t+3 u),\\
A^{3}_2 &=& \frac{-i d_1 \kappa}{6 L_1} (128 m_{c}^4 s_{w}+4 m_{c}^2 (s (8 s_{w}-6)-20 s_{w} t+36 s_{w} u+3 t-15 u)+\nonumber\\
&&t (16 s_{w} s-6 s+12 s_{w} t-36 s_{w} u-3 t+15 u)),\\
A^{3}_3 &=& \frac{i d_1 \kappa}{6 L_1} (64 m_{c}^4 (10 s_{w}-3)+4 m_{c}^2 (s (18-56 s_{w})-60 s_{w} t-52 s_{w} \nonumber\\
&&u+21 t+15 u)+ u (s (6-16 s_{w})+12 s_{w} (5 t+u)-3 (7 t+u))),\\
A^{3}_4 &=& \frac{i \kappa}{12 L_1} (2 d_1 (8 m_{c}^2 (8 s_{w}-3) (2s-4 m_{c}^2)+s^2 (6-16 s_{w})-3 s (4 s_{w}-1)\nonumber\\
  && (t-u)+2 (8 s_{w}-3) t u)+4(8 s_{w}-3) (t u-16 m_{c}^4) (d_2 +d_3) (t+u)),\\
A^{4}_1 &=& \frac{i d_1 m_{c}^2 \kappa}{3 L_1} (16 m_{c}^2 (8 s_{w}-3)+s (6-16 s_{w})-3 (4 s_{w}-1) (t-u)),\\
A^{4}_2 &=& \frac{i d_1 \kappa}{6 L_1}(128m_{c}^4s_{w}(8s_{w}-5)-4m_{c}^2(2(s(8s_{w}(8s_{w}-5)+3)+4s_{w}(8s_{w}-5)(t+3 u))-\nonumber\\
&&3(t-5u))+t(s(8 (5-8 s_{w}) s_{w}-6)+3 (8 s_{w} (8 s_{w}-5)+5) u-3 t))),\\
A^{4}_3 &=& \frac{id_1\kappa}{6L_1}(64m_{c}^4(2s_{w}(8s_{w}-5)+3)-4m_{c}^2(2(s(8s_{w}(8s_{w}-5)+9)+4s_{w}(8s_{w}-5)(3 t+u))\nonumber\\
&&+3 (7 t+5 u))+u (s (8 (5-8 s_{w}) s_{w}-6)+3 ((8 s_{w} (8 s_{w}-5)+7) t+u))),\\
A^{4}_4 &=& \frac{i \kappa}{12 L_1} (2 d_1 (4m_{c}^2(4s_{w}-1)(8s_{w}-3)(2s-4m_{c}^2)+s^2(8(5-8s_{w})\nonumber\\
&&s_{w}-6)+3s(u-t)+2(4s_{w}-1)(8s_{w}-3)t u)+(4s_{w}-1)(8s_{w}-3)(t u-16 m_{c}^4)\nonumber\\
&&(4(d_2+d_3)(t+u))).
\end{eqnarray}

\subsection{$ e^+(p_2) e^-(p_1) \to Z^0 \to J/\psi(q_1) + \chi_{cJ}(q_2)$}

There are twenty-two independent Lorentz structures for the process, which are
\begin{eqnarray}
B_{1} &=& \frac{i \varepsilon(p_{2}, \alpha, \beta, \sigma)}{m_{c}},\;\;
B_{2} = \frac{i \varepsilon(q_{1}, \alpha, \beta, \sigma)}{m_{c}},\;\;
B_{3} =\frac{i \varepsilon(q_{2}, \alpha, \beta, \sigma)}{m_{c}},\;\;
B_{4} = \frac{i p_{2}^{\sigma} \varepsilon(q_{1}, q_{2}, \alpha, \beta)}{m_{c}^3},\;\;
B_{5} = \frac{i p_{2}^{\beta} \varepsilon(q_{1}, q_{2}, \sigma, \alpha)}{m_{c}^3},\;\;\nonumber\\
B_{6} &=& \frac{i q_{1}^{\sigma} \varepsilon(p_{2}, q_{2}, \alpha, \beta)}{m_{c}^3},\;\;
B_{7} =  \frac{i q_{1}^{\beta} \varepsilon(p_{2}, q_{2}, \sigma, \alpha)}{m_{c}^3},\;\;
B_{8} = \frac{i q_{1}^{\beta } q_{1}^{\sigma} \varepsilon(p_{2}, q_{1}, q_{2}, \alpha)}{m_{c}^5},\;\;
B_{9} = \frac{ p_{2}^{\alpha } q_{1}^{\beta } q_{1}^{\sigma }}{m_{c}^3},\;\;\nonumber\\
B_{10} &=& \frac{i q_{2}^{\alpha} \varepsilon(p_{2}, q_{1}, \sigma, \beta)}{m_{c}^3},\;\;
B_{11} = \frac{i q_{2}^{\alpha} \varepsilon(p_{2}, q_{2}, \sigma, \beta)}{m_{c}^3},\;\;
B_{12} = \frac{i q_{2}^{\alpha} \varepsilon(q_{1}, q_{2}, \sigma, \beta)}{m_{c}^3},\;\;
B_{13} = \frac{i q_{2}^{\alpha } q_{1}^{\sigma} \varepsilon(p_{2}, q_{1}, q_{2}, \beta)}{m_{c}^5},\;\;\nonumber\\
B_{14} &=& \frac{ q_{2}^{\alpha } p_{2}^{\beta } q_{1}^{\sigma }}{m_{c}^3},\;\;
B_{15} = \frac{ q_{2}^{\alpha } q_{1}^{\beta } q_{1}^{\sigma }}{m_{c}^3},\;\;
B_{16} = \frac{i g^{\alpha \sigma } \varepsilon(p_{2}, q_{1}, q_{2}, \beta)}{m_{c}^3},\;\;
B_{17} = \frac{i g^{\beta \sigma } \varepsilon(p_{2}, q_{1}, q_{2}, \alpha)}{m_{c}^3},\;\;\nonumber\\
B_{18} &=& \frac{p_{2}^{\alpha } g^{\beta \sigma }}{m_{c}},\;\;
B_{19} = \frac{q_{2}^{\alpha } g^{\beta \sigma }}{m_{c}},\;\;
B_{20} = \frac{i g^{\alpha \beta } \varepsilon(p_{2}, q_{1}, q_{2}, \sigma)}{m_{c}^3},\;\;
B_{21} = \frac{i g^{\alpha \beta } p_{2}^{\sigma }}{m_{c}},\;\;
B_{22} = \frac{i g^{\alpha \beta } q_{1}^{\sigma }}{m_{c}},\;\;
\end{eqnarray}
whose non-zero coefficients $A^{3,4}_j$ ($j=1,2,....,22$) are
\begin{eqnarray}
A^{3}_1 &=& \frac{d_1 \kappa }{L_1 m_c}(t-u) (4 m_c^2+t+u),\;\;
A^{3}_2 = \frac{d_1 \kappa}{L_1 m_c} (16 m_c^4-4 m_c^2 t+u (u-t)),\\
A^{3}_3 &=& \frac{d_1 \kappa}{L_1 m_c} (4 m_c^2 u-t^2),\;\;
A^{3}_4 = A^{3}_5 = A^{3}_{10} =\frac{2 d_1 \kappa}{L_1 m_c} (u-t),\;\;
A^{3}_6 = A^{3}_7 = A^{3}_{16} = -A^{3}_4,\\
A^{3}_8 &=& -A^{3}_{13} = -\frac{16 m_c \kappa}{3 L_1} (4 s_{w}-1) (8 s_{w}-3) (d_{2}+d_{3}),\;\;
A^{3}_{9} = -A^{3}_{14} = -\frac{8 m_c \kappa}{3 L_1} (8 s_{w}-3) (d_{2}+d_{3}) (t-u),\\
A^{3}_{11} &=& -\frac{1}{2}A^{3}_4,\;\;
A^{3}_{12} = \frac{d_1 \kappa}{L_1 m_c} (4 m_c^2-2 t+u),\;\;
A^{3}_{15} = -\frac{8 m_c \kappa}{3 L_1} (8 s_{w}-3) (d_{2}+d_{3}) s,\\
A^{3}_{17} &=& \frac{4 d_1 \kappa}{3 L_1 m_c} (4 s_{w}-1) (8 s_{w}-3) s,\;\;
A^{3}_{18} = \frac{2 d_1 \kappa}{3 L_1 m_c} (8 s_{w}-3) (t-u) s,\\
A^{3}_{19} &=& \frac{2 d_1 \kappa(8 s_{w}-3)}{3 L_1 m_c}  (64 m_c^4-4 m_c^2 (3 t+u)+t (t-u)),\;\;
A^{3}_{20} = -\frac{2 d_1 \kappa}{3 L_1 m_c} (8 m_c^2 (4 s_{w}-1) (8 s_{w}-3)+3 (t-u)),\\
A^{3}_{21} &=& -\frac{4 m_c^2}{s} A^{3}_{18},\;\;
A^{3}_{22} = -\frac{8 m_c \kappa}{3 L_1} (8 s_{w}-3) (d_1 (u-4 m_c^2)+(d_{2}+d_{3}) (16 m_c^4-t u))  , \\
A^{4}_1 &=& (4 s_{w}-1) A^{3}_1,\;\;
A^{4}_2 = (4 s_{w}-1) A^{3}_2,\;\;
A^{4}_3 = (4 s_{w}-1) A^{3}_3,\;\;
A^{4}_4 = A^{4}_5 = A^{4}_{10} = (4 s_{w}-1) A^{3}_4,\\
A^{4}_6 &=& A^{4}_7 = A^{4}_{16} = -A^{4}_4,\;\;
A^{4}_8 = -A^{4}_{13} = \frac{A^{3}_8}{(4 s_{w}-1)},\;\;
A^{4}_{9} = -A^{4}_{14} = (4 s_{w}-1) A^{3}_9,\;\;
A^{4}_{11} = -\frac{1}{2}A^{4}_4,\\
A^{4}_{12} &=& (4 s_{w}-1) A^{3}_{12},\;\;
A^{4}_{15} = (4 s_{w}-1) A^{3}_{15},\;\;
A^{4}_{17} = \frac{A^{3}_{17}}{(4 s_{w}-1)},\;\;
A^{4}_{18} = (4 s_{w}-1) A^{3}_{18},\;\;
A^4_{19} = (4 s_{w}-1) A^{3}_{19},\\
A^{4}_{20} &=& -\frac{2 d_{1} \kappa }{3 L_1 m_c}(8 m_c^2 (8 s_{w}-3)+3 (4 s_{w}-1) (t-u)),\;\;
A^{4}_{21} = (4 s_{w}-1) A^{3}_{21},\;\;
A^{4}_{22} = (4 s_{w}-1) A^{3}_{22}.
\end{eqnarray}

Taking the property of $\varepsilon^{J}_{\alpha\beta}$, so practically we can safely set the coefficients to be zero:
\begin{equation}
\begin{array}{c}
A^i_j(|(c\bar{c})_{\bf 1}[^3P_0]\rangle)=0 \;\;\;\;\;\;\;\;{\rm for}\;i=(1-4), j=(1,2,3,10,11,12)
\end{array}
\end{equation}
\begin{equation}
\begin{array}{c}
A^i_j(|(c\bar{c})_{\bf 1}[^3P_1]\rangle)=0 \;\;\;\;\;\;\;\;{\rm for}\;i=(1-4), j=(8,9,15,17,18,19)
\end{array}
\end{equation}
\begin{equation}
\begin{array}{c}
A^i_j(|(c\bar{c})_{\bf 1}[^3P_2]\rangle)=0 \;\;\;\;\;\;\;\;{\rm for}\;i=(1-4), j=(1,2,3,10,11,12,17,18,19) .
\end{array}
\end{equation}

\section{The Lorentz structures and their coefficients for $\bm{e^+(p_{2})+e^-(p_{1})\to \gamma^*\to H_{1} (c\bar{c})(q_{1})+H_{1}(c\bar{c})(q_{2})}$}

In this section, we present the Lorentz structures and their nonzero coefficients for the process $e^+(p_{2})+e^-(p_{1})\to \gamma^*\to H_{1} (c\bar{c})(q_{1})+H_{1}(c\bar{c})(q_{2})$, where $n$ stands for the $|(c\bar{c})_{\bf 1}[^1S_0]\rangle$, $|(c\bar{c})_{\bf 1}[^3S_1]\rangle$, $|(c\bar{c})_{\bf 1}[^1P_1]\rangle$, and $|(c\bar{c})_{\bf 1}[^3P_J]\rangle(J=1,2,3)$. For the reason that we make a proximation $m_{e}=0$, the coefficients $A^{1,2}_j=0$. Some short notations appeared in the coefficients are defined as followings,
\begin{eqnarray}
s_{w}&=&{\sin ^2}{\theta _w},\;\;
L_{1}=\frac{{\cal C}}{\sqrt{2s}},\;\;
\kappa =\frac{2 i}{\sqrt{s} \sqrt{t u-16 m_{c}^4}},\;\;
d_{1}= \frac{8}{s^3 },\;\;
d_{2}=\frac{16}{s^4},\;\;
d_{3}= 2d_{2}.\nonumber
\end{eqnarray}
where, $s$, $t$, and $u$ are Mandelstam variables.

\subsection{S-wave + S-wave state: $ e^+e^- \to \gamma^*\to J/\psi+\eta_{c}$}

There are four Lorentz structures for the process, which are defined as,
\begin{eqnarray}
B_{1}&=&\frac{i}{m_{c}^3} \varepsilon(\alpha, p_1, p_2, q_1),\;\;
B_{2}=\frac{p_{1}^{\alpha }}{m_c},\;\;\nonumber
B_{3}=\frac{p_{2}^{\alpha }}{m_c},\;\;
B_{4}=\frac{q_{1}^{\alpha }}{m_c},\;\;\nonumber
\end{eqnarray}
whose non-zero coefficients $A^{3,4}_j$ ($j=1,2,3,4$) are
\begin{eqnarray}
A^3_2 = \frac{2 i m_{c} \kappa}{L_{1}} (t+4 m_{c}^2),\;\;
A^3_3 = -A^{3}_2|_{t\leftrightarrow u},\;\;
A^3_4 = -\frac{2 i m_{c} \kappa s}{L_{1}} (t-u),\;\;
A^4_1 = -\frac{2 m_{c}^2}{s} A^3_4.
\end{eqnarray}

\subsection{S-wave + P-wave state: $ e^+e^- \to \gamma^*\to \eta_{c}+h_{c}$}

There are four Lorentz structures for the process, which are defined as,
\begin{eqnarray}
B_{1}&=&\frac{i}{m_{c}^3} \varepsilon(\alpha, p_1, p_2, q_1),\;\;
B_{2}=\frac{p_{1}^{\alpha }}{m_c},\;\;\nonumber
B_{3}=\frac{p_{2}^{\alpha }}{m_c},\;\;
B_{4}=\frac{q_{1}^{\alpha }}{m_c},\;\;\nonumber
\end{eqnarray}
whose non-zero coefficients $A^{3,4}_j$ ($j=1,2,3,4$) are
\begin{eqnarray}
A^3_1 &=& \frac{2 i L_{1} m_{c}^2 \kappa}{L_{1}} (t+u),\;\;
A^4_2 = \frac{i d_{1} \kappa}{L_{1}} (16 m_{c}^4-4 m_{c}^2 (2 s+t+3 u)-t (s-3 u)),\\
A^4_3 &=& A^4_2|_{t\leftrightarrow u},\;\;
A^4_4 = \frac{i \kappa}{4 L_{1}} (4 d_{1} (16 m_{c}^4+4 m_{c}^2 (s-t-u)-s^2+t u)+4(t u-16m_{c}^4) (d_2+d_3)(t+u)).
\end{eqnarray}

\subsection{S-wave + P-wave state: $ e^+e^- \to \gamma^*\to J/\psi+\chi_{cJ}$}

There are four Lorentz structures for the process, which are defined as,
\begin{eqnarray}
B_{1}&=& -\frac{i q_{1}^{\beta } q_{1}^{\sigma } \varepsilon(\alpha, p_2, q_1, q_2)}{m_{c}^5},\;\;
B_{2}= \frac{i q_{1}^{\sigma } q_{2}^{\alpha } \varepsilon(\beta, p_2, q_1, q_2)}{m_{c}^5},\;\;
B_{3}= \frac{i g^{\beta \sigma } \varepsilon(\alpha, p_2, q_1, q_2)}{m_{c}^3},\;\;\nonumber\\
B_{4}&=& \frac{i g^{\alpha \beta } \varepsilon(\sigma, p_2, q_1, q_2)}{m_{c}^3},\;\;
B_{5}= \frac{ q_{1}^{\sigma } g^{\alpha \beta }}{m_{c}},\;\;
B_{6}= \frac{ p_{2}^{\sigma } g^{\alpha \beta }}{m_{c}},\;\;
B_{7}= \frac{ p_{2}^{\alpha } g^{\beta \sigma }}{m_{c}},\;\;\nonumber\\
B_{8}&=& \frac{ q_{2}^{\alpha } g^{\beta \sigma }}{m_{c}},\;\;
B_{9}=\frac{ p_{2}^{\alpha } q_{1}^{\beta } q_{1}^{\sigma }}{m_{c}^3},\;\;
B_{10}= \frac{ p_{2}^{\beta } q_{1}^{\sigma } q_{2}^{\alpha }}{m_{c}^3},\;\;
B_{11}= \frac{ q_{1}^{\beta } q_{1}^{\sigma } q_{2}^{\alpha }}{m_{c}^3},
\end{eqnarray}
whose coefficients $A^{3,4}_j$ ($j=1,2,3,4$) are
\begin{eqnarray}
A^3_1 &=& \frac{16 i m_{c}^6 \kappa}{L_{1}} (d_{2}+d_{3}),\;\;
A^3_2 = -A^3_{1},\;\;
A^3_3 = -\frac{4 i d_{1} m_{c}^4 \kappa s}{L_{1}},\;\;
A^3_4 = -\frac{4 m_{c}^2}{s}A^3_3,\\
A^4_5 &=& \frac{8 i m_{c}^4 \kappa}{L_{1}} ( d_{1} (u-4 m_{c}^2)+ d_{2} (16 m_{c}^4-ut)+d_{3} (8 m_{c}^4-ut)),\;\;
A^4_6 = \frac{8 i d_{1} m_{c}^2 \kappa}{L_{1}} (t-u),\;\;
A^4_7 = -\frac{s A^3_6}{2 m_{c}^2},\\
A^4_8 &=& -\frac{2 i d_{1} \kappa}{L_{1}} (64 m_{c}^4-4 m_{c}^2 (5t+u)+t (s+2 t)),\;\;
A^4_9 = -\frac{8 i m_{c}^4 \kappa}{L_{1}} (d_{2}+d_{3}) (u- t),\;\;
A^4_{10} = -A^4_{9},\;\;
A^4_{11} = \frac{s A^3_1}{2 m_{c}^2}.
\end{eqnarray}
We can safely set the coefficients before them to be zero with the property of $\varepsilon^{J}_{\alpha\beta}$:
\begin{equation}
\begin{array}{c}
A^i_j(|(c\bar{c})_{\bf 1}[^3P_1]\rangle)=0 \;\;\;\;\;\;\;\;{\rm for}\;i=(1-4),\; j=(1,3,7,8,9,11)
\end{array}
\end{equation}
\begin{equation}
\begin{array}{c}
A^i_j(|(c\bar{c})_{\bf 1}[^3P_2]\rangle)=0 \;\;\;\;\;\;\;\;{\rm for}\;i=(1-4),\;
j=(3,7,8) .
\end{array}
\end{equation}

\section{The Lorentz structures and their coefficients for $\bm{e^+(p_{2})+e^-(p_{1})\to \gamma^*\to (c\bar{c})[n](q_{1})+\gamma(q_{2})}$}

In this section, we present the Lorentz structures and their nonzero coefficients for the process $e^+(p_{2})+e^-(p_{1})\to \gamma^* \to (c\bar{c})[n](q_{1})+\gamma(q_{2})$, where $n$ stands for the $|(c\bar{c})_{\bf 1}[^1S_0]\rangle$, $|(c\bar{c})_{\bf 1}[^3S_1]\rangle$, $|(c\bar{c})_{\bf 1}[^1P_1]\rangle$, and $|(c\bar{c})_{\bf 1}[^3P_J]\rangle(J=1,2,3)$. For the reason that we make a proximation $m_{e}=0$, the coefficients $A^{1,2}_j=0$. Some short notations appeared in the coefficients are defined as followings,
\begin{eqnarray}
s_{w}&=&{\sin ^2}{\theta _w},\;\;
L_{1}=\frac{{\cal C}}{\sqrt{2s}},\;\;
\kappa =\frac{2 i}{ \sqrt{s t u}},\;\;
d_{1}=-\frac{2}{s (4 m_{c}^2-s)},\;\;
d_{2}=2 d_{1}.\nonumber
\end{eqnarray}
where, $s$, $t$, and $u$ are Mandelstam variables.

\subsection{Spin-singlet S-wave state: $|(c\bar{c})[^1S_0]\rangle$}

There are four Lorentz structures for the process, which are defined as,
\begin{eqnarray}
B_{1}&=&\frac{i}{m_{c}^3} \varepsilon(\nu, p_1, p_2, q_1),\;\;
B_{2}=\frac{p_{1}^{\nu }}{m_c},\;\;
B_{3}=\frac{p_{2}^{\nu }}{m_c},\;\;
B_{4}=\frac{q_{1}^{\nu }}{m_c},\;\;
\end{eqnarray}
whose coefficients $A^{3,4}_j$ ($j=1,2,3,4$) are
\begin{eqnarray}
A^3_2 &=& \frac{i \sqrt{m_{c}} \kappa}{\sqrt{2} L_{1}} (16 m_{c}^4-4 m_{c}^2 (s+2 t)+t(t+u)),\\
A^3_3 &=& A^3_2|_{t\leftrightarrow u},\;\;
A^3_4 = \frac{i \sqrt{m_{c}} \kappa s}{\sqrt{2} L_{1}} (t-u),\;\;
A^4_1 =  \frac{2 m_{c}^2}{s}A^3_4.
\end{eqnarray}

\subsection{Spin-triplet P-wave state: $|(c\bar{c})[^3P_J]\rangle$}

There fifteen Lorentz structures for the process, which are defined as,
\begin{eqnarray}
B_{1}&=&\frac{i}{m_{c}^3} g^{\beta \nu } \varepsilon(p_2, q_1, q_2, \alpha),\;\;
B_{2}=\frac{i}{m_{c}^3} g^{\alpha \nu } \varepsilon(p_2, q_1, q_2, \beta),\;\;
B_{3}=\frac{i}{m_{c}^5} q_{1}^{\nu } q_{2}^{\beta } \varepsilon(p_2, q_1, q_2, \alpha),\nonumber\\
B_{4}&=&\frac{i}{m_{c}^5} q_{2}^{\nu } q_{2}^{\beta } \varepsilon(p_2, q_1, q_2, \alpha),\;\;
B_{5}=\frac{i}{m_{c}^5} q_{1}^{\nu } q_{2}^{\alpha} \varepsilon(p_2, q_1, q_2, \beta),\;\;
B_{6}=\frac{i}{m_{c}^5}  q_{2}^{\alpha } q_{2}^{\beta } \varepsilon(\nu, p_2, q_1, q_2),\nonumber\\
B_{7}&=&\frac{1}{m_{c}} g^{\alpha \nu } p_{2}^{\beta },\;\;
B_{8}=\frac{1}{m_{c}} g^{\alpha \nu } q_{2}^{\beta },\;\;
B_{9}=\frac{1}{m_{c}} g^{\beta \nu } p_{2}^{\alpha},\;\;
B_{10}=\frac{1}{m_{c}} g^{\beta \nu } q_{2}^{\alpha},\;\;
B_{11}=\frac{1}{m_{c}^3}  q_1^{\nu} p_{2}^{\alpha } q_{2}^{\beta },\;\;\nonumber\\
B_{12}&=&\frac{1}{m_{c}^3}  q_1^{\nu} q_{2}^{\alpha } p_{2}^{\beta },\;\;
B_{13}=\frac{1}{m_{c}^3}  q_2^{\nu} p_{2}^{\alpha } q_{2}^{\beta },\;\;
B_{14}=\frac{1}{m_{c}^3}  p_2^{\nu} q_{2}^{\alpha } q_{2}^{\beta },\;\;
B_{15}=\frac{1}{m_{c}^3}   q_1^{\nu} q_{2}^{\alpha } q_{2}^{\beta }.
\end{eqnarray}
The non-zero coefficients $A^{3}_j$ and $A^{4}_j$ that are the same for all three $P$-waves:
\begin{eqnarray}
A^{3}_{1} &=&-\frac{i \sqrt{2} d_{1} m_{c}^{3/2} \kappa}{L_{1}} (4 m_{c}^2+s),\;\;
A^{3}_{2} = A^{3}_{1}|_{s\leftrightarrow -s},\;\;
A^{3}_{3} = \frac{2 i \sqrt{2} m_{c}^{7/2} \kappa}{L_{1}} (d_{1}+4 d_{2} m_{c}^2),\\
A^{3}_{4} &=& A^{3}_{6}= \frac{8 i \sqrt{2} d_{2} m_{c}^{11/2} \kappa}{L_{1}},\;\;
A^{3}_{5} = -\frac{d_{1} A^{3}_{4}}{4 d_{2} m_{c}^2},\;\;
A^{4}_{7} = \frac{i d_{1} \kappa (u^2-t^2) }{\sqrt{2} L_{1} \sqrt{m_{c}}},\\
A^{4}_{8} &=& \frac{i \kappa}{\sqrt{2} L_{1} \sqrt{m_{c}}} (d_{1} (u+s) (u- t)-8 m_{c}^2 u (d_{1}+d_{2} t)),\;\;
A^{4}_{9} = \frac{i d_{1} \kappa}{\sqrt{2} L_{1} \sqrt{m_{c}}} (4 m_{c}^2+s) (u- t),\\
A^{4}_{10} &=& -\frac{i d_{1} \kappa}{\sqrt{2} L_{1} \sqrt{m_{c}}} (48 m_{c}^4-4 m_{c}^2 (s+5 t+2 u)+t (s+2 t)),\;\;
A^{4}_{11} = -\frac{i \sqrt{2} m_{c}^{3/2} \kappa}{L_{1}} (d_{1}+4 d_{2} m_{c}^2) (u - t),\\
A^{4}_{12} &=& \frac{i \sqrt{2} d_{1} m_{c}^{3/2} \kappa}{L_{1}} (u - t),\;\;
A^{4}_{13} = -A^{4}_{14}= -\frac{4 i \sqrt{2} d_{2} m_{c}^{7/2} \kappa}{L_{1}} (u - t),\;\;
A^{4}_{15} = \frac{4 i \sqrt{2} d_{2} m_{c}^{7/2} \kappa s}{L_{1}}.
\end{eqnarray}
Noticing the properties of polarization tensor $\varepsilon^{J}_{\alpha\beta}$, we have $A_{6}^{4}=A_{14}^{4}=A_{15}^{4}=0$ for the $^3P_1$ state.

\section{The Lorentz structures and their coefficients for $\bm{e^+(p_{2})+e^-(p_{1})\to Z^{0}\to (c\bar{c})[n](q_{1})+\gamma(q_{2})}$}

In this section, we present the Lorentz structures and their nonzero coefficients for the process $e^+(p_{2})+e^-(p_{1})\to Z^{0}\to (c\bar{c})[n](q_{1})+\gamma(q_{2})$, where $n$ stands for the $|(c\bar{c})_{\bf 1}[^1S_0]\rangle$, $|(c\bar{c})_{\bf 1}[^3S_1]\rangle$, $|(c\bar{c})_{\bf 1}[^1P_1]\rangle$, and $|(c\bar{c})_{\bf 1}[^3P_J]\rangle(J=1,2,3)$. For the reason that we make a proximation $m_{e}=0$, the coefficients $A^{1,2}_j=0$. Some short notations appeared in the coefficients are defined as followings,
\begin{eqnarray}
s_{w}&=&{\sin ^2}{\theta _w},\;\;
L_{1}=\frac{{\cal C}}{\sqrt{2s}},\;\;
\kappa =\frac{2}{\sqrt{s} \sqrt{t u-16 m_{c}^4}},\nonumber\\
d_{1}&=&\frac{1}{6 \sqrt{3} \sqrt{1-s_{w}} (\frac{s}{2}-2 m_{c}^2) \sqrt{\Gamma_z^2 m_{z}^2+(s-m_{z}^2)^2}},\nonumber\\
d_{2}&=&\frac{1}{6 \sqrt{3} \sqrt{1-s_{w}} (4 m_{c}^4-2 m_{c}^2 s+\frac{s^2}{4}) \sqrt{\Gamma_z^2 m_{z}^2+(s-m_{z}^2)^2}}.\nonumber
\end{eqnarray}
where, $s$, $t$, and $u$ are Mandelstam variables.

\subsection{Spin-singlet S-wave state: $|(c\bar{c})[^1S_0]\rangle$}

There are four Lorentz structures for $|(c\bar{c})[^1S_0]\rangle$ through $Z^0$-propagator, which are defined as,
\begin{eqnarray}
B_{1}&=&\frac{i}{m_{c}^3} \varepsilon(\nu, p_1, p_2, q_1),\;\;
B_{2}=\frac{p_{1}^{\nu }}{m_c},\;\;\nonumber
B_{3}=\frac{p_{2}^{\nu }}{m_c},\;\;
B_{4}=\frac{q_{1}^{\nu }}{m_c},\;\;\nonumber
\end{eqnarray}
whose non-zero coefficients $A^{3,4}_j$ ($j=1,2,3,4,5$) are
\begin{eqnarray}
A^{3}_1 &=& -\frac{i \sqrt{2} m_{c}^{5/2} \kappa}{3 L_{1}} (8 s_{w}-3) (t-u),\;\;
A^{3}_2 = -\frac{i \sqrt{m_{c}} \kappa}{3 \sqrt{2} L_{1}} (4 s_{w}-1) (8 s_{w}-3) (16 m_{c}^4-4 m_{c}^2 (s+2 t)+t (t+u)),\\
A^{3}_3 &=& A^{3}_2|_{t\leftrightarrow u},\;\;
A^{3}_4 = \frac{s (4 s_{w}-1)}{2 m_{c}^2} A^{3}_1,\;\;
A^{4}_1 = (4 s_{w}-1) A^{3}_1,\;\;
A^{4}_2 = \frac{A^3_2}{4 s_w-1},\;\;
A^{4}_3 = A^{4}_2|_{t\leftrightarrow u},\;\;
A^{4}_4 = \frac{s}{2 m_{c}^2} A^{3}_1.
\end{eqnarray}

\subsection{Spin-triplet S-wave state: $|(c\bar{c})[^3S_1]\rangle$}

There seven Lorentz structures for the process, which are listed as followings:
\begin{eqnarray}
B_{1}&=&\frac{i}{m_{c}^2} \varepsilon(\alpha, \nu, p_{1}, q_{2}),\;\;
B_{2}=\frac{i}{m_{c}^2} \varepsilon(\alpha, \nu, p_{2}, q_{2}),\;\;
B_{3}=\frac{i}{m_{c}^2} \varepsilon(\alpha, \nu, q_{1}, q_{2}),\nonumber\\
B_{4}&=&\frac{1}{m_{c}^2} p_{1}^{\nu } p_{2}^{\alpha },\;\;
B_{5}=\frac{1}{m_{c}^2} p_{1}^{\alpha } p_{2}^{\nu },\;\;
B_{6}=\frac{1}{m_{c}^2} p_{1}^{\alpha } q_{1}^{\nu },\;\;
B_{7}=\frac{1}{m_{c}^2} p_{2}^{\alpha } q_{1}^{\nu },\;\;\nonumber
\end{eqnarray}
whose non-zero coefficients are:
\begin{eqnarray}
A^{3}_1 &=& \frac{i 2 \sqrt{2} m_{c}^{5/2} \kappa }{L_{1}}(s+u),\;\;
A^{3}_2 = A^{3}_1|_{t\leftrightarrow u},\;\;
A^{3}_3 =-\frac{i 2 \sqrt{2} m_{c}^{5/2} \kappa s}{L_{1}},\;\;
A^{3}_4 =\frac{i 2 \sqrt{2} m_{c}^{5/2} \kappa }{L_{1}}(4 s_{w}-1) (u+t),\\
A^{3}_5 &=& -A^{3}_4,\;\;
A^{3}_6 =\frac{i 2 \sqrt{2} m_{c}^{5/2} \kappa u}{L_{1}} (4 s_{w}-1),\;\;
A^{4}_7 = -A^{3}_6|_{t\leftrightarrow u},\;\;
A^{4}_1 =\frac{i 2 \sqrt{2} m_{c}^{5/2} \kappa}{L_{1}} (4 s_{w}-1) (s+u),\\
A^{4}_2 &=& A^{4}_1|_{t\leftrightarrow u},\;\;
A^{4}_3 =\frac{i 2 \sqrt{2} m_{c}^{5/2} \kappa s}{L_{1}} (1-4 s_{w}),\;\;
A^{4}_4 =\frac{i 2 \sqrt{2} m_{c}^{5/2} \kappa}{L_{1}} (u+t),\;\;
A^{4}_5 = -A^{4}_4,\\
A^{4}_6 &=&\frac{i 2 \sqrt{2} m_{c}^{5/2} \kappa u}{L_{1}},\;\;
A^{4}_7 = -A^{4}_6|_{t\leftrightarrow u}.
\end{eqnarray}

\subsection{Spin-singlet P-wave state: $|(c\bar{c})[^1P_1]\rangle$}

There eleven Lorentz structures for the process, which are defined as,
\begin{eqnarray}
B_{1}&=& g^{\alpha \nu },\;\;
B_{2}=\frac{1}{m_{c}^2} q_{2}^{\alpha } p_{1}^{\nu } ,
B_{3}=\frac{1}{m_{c}^2} q_{2}^{\alpha } p_{2}^{\nu } ,\;\;
B_{4}=\frac{1}{m_{c}^2} q_{2}^{\alpha } q_{1}^{\nu },\nonumber\\
B_{5}&=&\frac{1}{m_{c}^2} q_{2}^{\alpha } q_{2}^{\nu },\;\;
B_{6}=\frac{1}{m_{c}^2}  q_{1}^{\alpha } p_{1}^{\nu },\;\;
B_{7}=\frac{1}{m_{c}^2} q_{1}^{\alpha } p_{2}^{\nu },\;\;
B_{8}=\frac{1}{m_{c}^2} q_{1}^{\alpha } q_{1}^{\nu },\nonumber\\
B_{9}&=&\frac{i}{m_{c}^4} q_{1}^{\alpha } \varepsilon(\nu, p_{1}, p_{2}, q_{1}),\;\;
B_{10}=\frac{i}{m_{c}^4}  q_{2}^{\alpha } \varepsilon(\nu, p_{1}, p_{2}, q_{1}),\;\;
B_{11}=\frac{i}{m_{c}^6} q_{2}^{\alpha } q_{1}^{\nu } \varepsilon(p_{1}, p_{2}, q_{1}, q_{2}),
\end{eqnarray}
whose non-zero coefficients are,
\begin{eqnarray}
A^{3}_1 &=& \frac{2 i \sqrt{2} d_{1} \kappa  t u}{L_{1} \sqrt{m_{c}}},\;\;
A^{3}_2 = -\frac{i \sqrt{2} d_{2} \kappa  m_{c}^{3/2}}{L_{1}} (u+t) (s+u),\;\;
A^{3}_3 = A^{3}_2|_{t\leftrightarrow u},\;\;
A^{3}_4 = -\frac{i \sqrt{2} d_{2} \kappa  m_{c}^{3/2}}{L_{1}} ((t+u)^2-t-u),\\
A^{3}_5 &=&-\frac{d_{2} m_{c}^2}{d_{1}} A^{3}_1,\;\;
A^{3}_6 = \frac{d_{1} A^{3}_2}{d_{2} (u+t)},\;\;
A^{3}_7 = A^{3}_6|_{t\leftrightarrow u},\;\;
A^{3}_8 = \frac{m_{c}^2 s}{2 t u} A^{3}_1,\;\;
A^{3}_9 = \frac{2 i \sqrt{2} d_{1} \kappa  m_{c}^{7/2}}{L_{1}} (4 s_{w}-1),\\
A^{3}_{10} &=&\frac{d_{2}}{d_{1}} (u+t) A^{3}_9,\;\;
A^{3}_{11} = -\frac{2 d_{2} m_{c}^2}{d_{1}} A^{3}_9,\;\;
A^{4}_1 = (4 s_{w}-1)A^{3}_1,\;\;
A^{4}_2 =(4 s_{w}-1)A^{3}_2,\;\;
A^{4}_3 = A^{4}_2|_{t\leftrightarrow u},\\
A^{4}_4 &=& (4 s_{w}-1)A^{3}_4,\;\;
A^{4}_5 =(4 s_{w}-1)A^{3}_5,\;\;
A^{4}_6 = (4 s_{w}-1)A^{3}_6,\;\;
A^{4}_7 = A^{4}_6|_{t\leftrightarrow u},\;\;
A^{4}_8 = (4 s_{w}-1)A^{3}_8,\\
A^{4}_9 &=&\frac{A^{3}_9}{4 s_{w}-1},\;\;
A^{4}_{10} = \frac{A^{3}_{10}}{4 s_{w}-1}\;\;
A^{4}_{11} = \frac{A^{3}_{11}}{4 s_{w}-1}.
\end{eqnarray}

\subsection{Spin-triplet P-wave state: $|(c\bar{c})[^3P_J]\rangle$}

There twenty five Lorentz structures for the process, which are defined as,
\begin{eqnarray}
B_{1}&=&\frac{i}{m_{c}}\varepsilon(q_{1}, \nu, \alpha, \beta,),\;\;
B_{2}=\frac{i}{m_{c}}\varepsilon(q_{2}, \nu, \alpha, \beta,),\;\;
B_{3}=\frac{i}{m_{c}}\varepsilon(p_{2}, \nu, \alpha, \beta,),\nonumber\\
B_{4}&=& \frac{1}{m_{c}} g^{\alpha \nu } p_{2}^{\beta },\;\;
B_{5}= \frac{1}{m_{c}} g^{\beta \nu } p_{2}^{\alpha },\;\;
B_{6}= \frac{1}{m_{c}} g^{\alpha \nu } q_{2}^{\beta },\;\;
B_{7}=\frac{1}{m_{c}} g^{\beta \nu } q_{2}^{\alpha },\;\;
B_{8}=\frac{1}{m_{c}^3} q_{1}^{\nu } p_{2}^{\alpha } q_{2}^{\beta }, \nonumber\\
B_{9}&=&\frac{1}{m_{c}^3} q_{2}^{\nu } p_{2}^{\alpha } q_{2}^{\beta },\;\;
B_{10}= \frac{1}{m_{c}^3} p_{2}^{\nu } q_{2}^{\alpha } q_{2}^{\beta },\;\;
B_{11}=\frac{1}{m_{c}^3} q_{1}^{\nu } q_{2}^{\alpha } q_{2}^{\beta },\;\;
B_{12}=\frac{i}{m_{c}^3}  \varepsilon(p_{2}, q_{1}, q_{2}, \alpha) g^{\beta \nu },\nonumber\\
B_{13}&=&\frac{i}{m_{c}^3}  \varepsilon(p_{2}, q_{1}, q_{2}, \beta) g^{\alpha \nu },\;\;
B_{14}=\frac{i}{m_{c}^5} \varepsilon(p_{2}, q_{1}, q_{2}, \alpha) q_{1}^{\nu } q_{2}^{\beta},\;\;
B_{15}=\frac{i}{m_{c}^5} \varepsilon(p_{2}, q_{1}, q_{2}, \beta) q_{1}^{\nu } q_{2}^{\alpha},\nonumber\\
B_{16}&=&\frac{i}{m_{c}^5}  \varepsilon(\nu, p_{2}, q_{1}, q_{2}) q_{2}^{\alpha} q_{2}^{\beta},\;\;
B_{17}=\frac{i}{m_{c}^3} \varepsilon(p_{2}, q_{1}, \beta, \alpha) q_{1}^{\nu },\;\;
B_{18}=\frac{i}{m_{c}^3} \varepsilon(p_{2}, q_{2}, \beta, \alpha) q_{1}^{\nu },\nonumber\\
B_{19}&=&\frac{i}{m_{c}^3} \varepsilon(q_{1}, q_{2}, \beta, \alpha) q_{1}^{\nu },\;\;
B_{20}=\frac{i}{m_{c}^3}  \varepsilon(\nu, q_{1}, q_{2}, \alpha) p_{2}^{\beta },\;\;
B_{21}=\frac{i}{m_{c}^3}  \varepsilon(\nu, q_{1}, q_{2}, \beta) p_{2}^{\alpha },\nonumber\\
B_{22}&=&\frac{i}{m_{c}^3}  \varepsilon(\nu, p_{2}, q_{1}, \alpha) q_{2}^{\beta },\;\;
B_{23}=\frac{i}{m_{c}^3}  \varepsilon(\nu, p_{2}, q_{1}, \beta) q_{2}^{\alpha },\;\;
B_{24}=\frac{i}{m_{c}^3}  \varepsilon(p_{2}, q_{1}, q_{2}, \alpha) q_{2}^{\nu } q_{2}^{\beta },\nonumber\\
B_{25}&=&\frac{1}{m_{c}^3} q_{1}^{\nu } q_{2}^{\alpha } p_{2}^{\beta }.\nonumber
\end{eqnarray}

The non-zero coefficients $A^{3}_j$ and $A^{4}_j$ for all three $P$-waves:
\begin{eqnarray}
A^{3}_{1} &=& \frac{i d_{1} \kappa u(s-t)}{\sqrt{2} L_1 \sqrt{m_{c}}},\;\;
A^{3}_{2} = \frac{i d_{1} \kappa}{\sqrt{2} L_1 \sqrt{m_{c}}} (s^2+s u-2ut),\;\;
A^{3}_{3} = \frac{i d_{1} \kappa s}{\sqrt{2} L_1 \sqrt{m_{c}}} (u+t),\\
A^{3}_{4} &=& -\frac{i d_{1} \kappa}{3 \sqrt{2} L_1 \sqrt{m_{c}}} (8 s_{w}-3) (4 m_{c}^2-s) (u- t),\;\;
A^{3}_{5} = A^{3}_{4}|_{s\leftrightarrow -s},\\
A^{3}_{6} &=& -\frac{i \kappa}{3 \sqrt{2} L_1 \sqrt{m_{c}}} (8 s_{w}-3) (d_{1} (s+u) (u- t)-8 m_{c}^2 u (d_{1}+d_{2} t)),\\
A^{3}_{7} &=& \frac{i d_{1} \kappa}{3 \sqrt{2} L_1 \sqrt{m_{c}}} (8 s_{w}-3) (2s^2+u^2+st+3su-ut),\;\;
A^{3}_{8} = \frac{i \sqrt{2} m_{c}^{3/2} \kappa}{3 L_1} (8 s_{w}-3) (d_{1}+4 d_{2} m_{c}^2) (u- t),\\
A^{3}_{9} &=& \frac{4 i \sqrt{2} d_{2} m_{c}^{7/2} \kappa}{3 L_1} (8 s_{w}-3) (u-t),\;\;
A^{3}_{10} = -A^{3}_{9},\;\;
A^{3}_{11} = -\frac{s}{u-t} A^{3}_{9},\\
A^{3}_{12} &=& \frac{2 i \sqrt{2} d_{1} m_{c}^{3/2} \kappa}{3 L_1} (8 m_{c}^2 s_{w} (8 s_{w}-5)+s (2 s_{w} (8 s_{w}-5)+3)+3 t),\\
A^{3}_{13} &=& \frac{2 i \sqrt{2} d_{1} m_{c}^{3/2} \kappa}{3 L_1} ((2 s_{w} (8 s_{w}-5)+3) (t+u)-3 t),\;\;
A^{3}_{14} = \frac{2 m_{c}^2}{u-t} (4 s_{w}-1) A^{3}_{8},\;\;
A^{3}_{15} = \frac{A^{3}_{14}}{d_1 + 4 d_2 m_{c}^2},\\
A^{3}_{16} &=& -\frac{4 d_{2} m_{c}^2}{d_1} A^{3}_{15},\;\;
A^{3}_{17} = -\frac{i \sqrt{2} d_{1} m_{c}^{3/2} \kappa}{2 L_1} (u- t),\;\;
A^{3}_{18} = 2 A^{3}_{17},\;\;
A^{3}_{19} = -\frac{i d_{1} m_{c}^{3/2} \kappa }{\sqrt{2} L_1}(s- u),\\
A^{3}_{20} &=& -A^{3}_{21} = A^{3}_{22} = -A^{3}_{23} = -A^{3}_{18},\;\;
A^{3}_{24} = A^{3}_{16},\;\;
A^{3}_{25} = -\frac{d_{1} A^{3}_{9}}{4 m_{c}^2 d_2},\\
A^{4}_{1} &=& (4 s_{w}-1) A^{3}_{1},\;\;
A^{4}_{2} = (4 s_{w}-1) A^{3}_{2},\;\;
A^{4}_{3} = \frac{i d_{1} \kappa s}{\sqrt{2} L_1 \sqrt{m_{c}}} (4 s_{w}-1) (t-u),\;\;
A^{4}_{4} = (4 s_{w}-1) A^{3}_{4},\\
A^{4}_{5} &=& A^{4}_{4}|_{s\leftrightarrow -s},\;\;
A^{4}_{6} = (4 s_{w}-1) A^{3}_{6},\;\;
A^{4}_{7} = (4 s_{w}-1) A^{3}_{7},\;\;
A^{4}_{8} = (4 s_{w}-1) A^{3}_{8},\;\;
A^{4}_{9} = (4 s_{w}-1) A^{3}_{9},\\
A^{4}_{10} &=& -A^{4}_{9},\;\;
A^{4}_{11} = (4 s_{w}-1) A^{3}_{11},\;\;
A^{4}_{12} = -\frac{2 i \sqrt{2} d_{1} m_{c}^{3/2} \kappa}{3 L_1} (8 m_{c}^2 s_{w}+s (3-10 s_{w})+3 (1-4 s_{w}) t),\\
A^{4}_{13} &=& \frac{2 i \sqrt{2} d_{1} m_{c}^{3/2} \kappa}{3 L_1} ((10 s_{w}-3) (u+t)+3 (1-4 s_{w}) t),\;\;
A^{4}_{14} = \frac{A^{3}_{14}}{4 s_{w}-1},\;\;
A^{4}_{15} = \frac{A^{3}_{15}}{4 s_{w}-1},\;\;
A^{4}_{16} = \frac{A^{3}_{16}}{4 s_{w}-1},\\
A^{4}_{17} &=& (4 s_{w}-1) A^{3}_{17},\;\;
A^{4}_{18} = 2 A^{4}_{17},\;\;
A^{4}_{19} = (4 s_{w}-1) A^{3}_{19},\;\;
A^{4}_{20} = -A^{4}_{21} = A^{4}_{22} = -A^{4}_{23} = -A^{4}_{18},\\
A^{4}_{24} &=& \frac{A^{3}_{24}}{4 s_{w}-1},\;\;
A^{4}_{25} = (4 s_{w}-1) A^{3}_{25}.
\end{eqnarray}
For the reason of $\varepsilon^J_{\alpha\beta}$, we have:
\begin{equation}
\begin{array}{c}
A^i_j(|(c\bar{c})_{\bf 1}[^3P_0]\rangle)=0 \;\;\;\;\;\;\;\;{\rm for}\;i=(1-4), j=(1,2,3,17,18,19)
\end{array}
\end{equation}
\begin{equation}
\begin{array}{c}
A^i_j(|(c\bar{c})_{\bf 1}[^3P_1]\rangle)=0 \;\;\;\;\;\;\;\;{\rm for}\;i=(1-4), j=(10,11,16)
\end{array}
\end{equation}
\begin{equation}
\begin{array}{c}
A^i_j(|(c\bar{c})_{\bf 1}[^3P_2]\rangle)=0 \;\;\;\;\;\;\;\;{\rm for}\;i=(1-4), j=(1,2,3,17,18,19) .
\end{array}
\end{equation}

\end{widetext}

\end{document}